\documentclass[onecolumn,preprintnumbers,amsmath,amssymb,notitlepage]{revtex4}
\usepackage{graphicx}
\usepackage{epsfig}
\usepackage{amsmath}
\usepackage{amssymb}
\usepackage{dcolumn}

\begin{document}

\title{Canonical Quantum Gravity on Noncommutative Spacetime}

\author{Martin Kober}
\email{MartinKober@t-online.de}

\affiliation{Kettenhofweg 121, 60325 Frankfurt am Main, Germany}

\begin{abstract}
In this paper canonical quantum gravity on noncommutative space-time is considered. The corresponding generalized
classical theory is formulated by using the moyal star product, which enables the representation of the field
quantities depending on noncommuting coordinates by generalized quantities depending on usual coordinates. But not
only the classical theory has to be generalized in analogy to other field theories. Besides, the necessity arises
to replace the commutator between the gravitational field operator and its canonical conjugated quantity by a
corresponding generalized expression on noncommutative space-time. Accordingly the transition to the quantum theory
has also to be performed in a generalized way and leads to extended representations of the quantum theoretical
operators. If the generalized representations of the operators are inserted to the generalized constraints, one
obtains the corresponding generalized quantum constraints including the Hamiltonian constraint as dynamical
constraint. After considering quantum geometrodynamics under incorporation of a coupling to matter fields, the
theory is transferred to the Ashtekar formalism. The holonomy representation of the gravitational field as it is
used in loop quantum gravity opens the possibility to calculate the corresponding generalized area operator. 
\end{abstract}

\maketitle

\section{Introduction}

A decisive problem in contemporary theoretical physics is the search for a quantum theory of general relativity.
One of the most important approaches to formulate such a theory is the canonical quantization of gravity.
In particular in its special manifestation as loop quantum gravity, which foundations have been developed
in \cite{Rovelli:1989za}, explored in \cite{Rovelli:1991zi},\cite{Ashtekar:1992tb},\cite{Rovelli:1993bm},\cite{Rovelli:1995ac} for example and which has been reviewed in \cite{Nicolai:2005mc},\cite{Thiemann:2006cf},\cite{Rovelli:2010zz},\cite{Rovelli:2010vv} for example,
and which is based on the formulation of canonical gravity given in \cite{Ashtekar:1986yd},\cite{Ashtekar:1987gu},\cite{Ashtekar:1989ju}
it is considered as a very promising candidate for a quantum theoretical formulation of general relativity.
Another important concept in the context of quantum gravity is noncommutative geometry, which goes hand in hand
with the introduction of a minimal length scale appearing already at the classical level, which means before the
quantization of the fields, and was first suggested in \cite{Snyder:1946qz}. Accordingly, the concept of
noncommutative geometry could lead to the possibility to circumvent the necessity of renormalization in
quantum field theory and perhaps also in the usual field theoretic description of quantum gravity.
The idea of noncommutative geometry can in principle also be extended to noncommuting momenta leading to
the possibility to cure infrared divergencies besides ultraviolet divergencies \cite{Kober:2010um}. 
If a formulation of quantum field theory, which is really satisfying, should only be possible on a noncommutative space-time, then also any approach to quantum gravity would have to be formulated on noncommutative space-time. Since the canonical
quantization of general relativity is a very natural way to try to obtain a quantum description of general relativity,
under this assumption it would indeed seem to be necessary to formulate canonical quantum gravity on noncommutative space-time. Various theories of gravity in the context of noncommutative geometry have amongst others been studied in
\cite{Chamseddine:1992yx},\cite{Madore:1993br},\cite{Hawkins:1996uu},\cite{Avramidi:2003mk},\cite{GarciaCompean:2003pm},\cite{Calmet:2005qm},\cite{Calmet:2006iz},\cite{Aschieri:2005yw},\cite{Aschieri:2005zs},\cite{Harikumar:2006xf},\cite{Vassilevich:2009cb},\cite{Kober:2011rb}
and cosmology on noncommutative space-time has been explored in \cite{Barbosa:2004kp}.
Even certain ideas referring to quantum gravity have been explored with respect to noncommutative geometry, see \cite{Moffat:2000fv},\cite{Moffat:2000gr},\cite{Acatrinei:2002sb},\cite{Vassilevich:2004ym},\cite{Rosenbaum:2006yu},\cite{Faizal:2011wm},\cite{Faizal:2012kc},\cite{Faizal:2013ioa} for example.
However, canonical quantum gravity on a space-time with noncommutative coordinates as it is mentioned above has not been
considered so far. It is the aim of this paper to formulate such a theory. By using the moyal star product, products of fields depending on noncommuting coordinates can be mapped to products of fields depending on usual coordinates.
Accordingly it is possible to represent a field theory on noncommutative space-time as generalized field theory on usual
space-time. To obtain the quantum theoretical setting of this theory, a generalized expression for the commutator of two
operators on noncommutative space-time has to be derived, since an arbitrary commutator between quantities depending on noncommuting coordinates has to be extended as well, if it is reexpressed by quantities depending on commuting coordinates.
This leads to a generalized quantization condition for the canonical quantization of field theories, which has to be used to quantize the generalized canonical theory of gravity. In accordance with that, to obtain the canonical quantum theory of gravity, it is not only necessary to transfer the usual canonical formulation of classical general relativity to a noncommutative space-time, but it is also necessary to extend the quantization condition. The modification of the quantization procedure implies generalized operators describing the corresponding kinematics of quantum gravity. Because of this and because of the extension of the classical theory, one obtains generalized quantum constraints and thus also generalized quantum dynamics.
To formulate the generalized quantum constraints, the three metric representation of the generalized operators describing
the gravitational field is used.
Such a generalization of the quantization principle in canonical quantum gravity as it is implied by the noncommutativity
of space-time presupposed in this paper, has in other manifestations been considered in \cite{Kober:2011uj},\cite{Kober:2015bkv},\cite{Kober:2014xxa},\cite{Majumder:2011ad},\cite{Majumder:2011bv}.
A canonical noncommutativity algebra between the components of the gravitational field and the corresponding 
generalization of general relativity has been considered in \cite{Kober:2011am}.
After formulating the generalized canonical quantum theory of gravity by referring to quantum geometrodynamics,
a transition to Ashtekars formalism is performed in this paper. The incorporation of matter is considered as well.
Based on the corresponding holonomy representation of the gravitational field as it is used in loop quantum gravity,
the generalized area operator can be determined, what is done in the last section.

\section{The Noncommutative Geometry Scenario}

As basic assumption, a space-time with coordinates fulfilling a canonical noncommutativity algebra is postulated.
Accordingly the coordinates become operators, $x^\mu \rightarrow \hat x^\mu$, and they obey a commutation relation
of the following form:

\begin{equation}
\left[\hat x^\mu,\hat x^\nu \right]=i\theta^{\mu\nu},
\end{equation}
where $\theta^{\mu\nu}$ is an antisymmetric tensor of second order rather than a constant what
maintains Lorentz invariance. In accordance with the splitting of space-time into a spacelike
hypersurface and a time coordinate, which will be presupposed in the next section, it will be
made the assumption that only the components referring to the spacelike hypersurface at every
point obey nontrivial commutation relations. This means that the noncommutativity algebra
takes the following special shape, where only the spacelike components are incorporated:

\begin{equation}
\left[\hat x^m,\hat x^n \right]=i\theta^{mn},\quad {\rm with} \quad
\theta=\left(\begin{matrix}
0 & \theta_{12} & \theta_{13}\\
-\theta_{12} & 0 & \theta_{23}\\
-\theta_{13} & -\theta_{23} & 0
\end{matrix}\right).
\label{algebra}
\end{equation}
The noncommutative geometry induced by the algebra ($\ref{algebra}$) of course also leads to generalized properties
of field theories formulated on a space-time with such coordinates. It is possible to represent a field theory
according to ($\ref{algebra}$) as generalized field theory on usual space-time. This can be done by mapping
the appearing products of fields in such a theory, which depend on noncommuting coordinates, to generalized
products of such fields depending on usual coordinates.
The product of two arbitrary fields, $\phi(\hat x)$ and $\psi(\hat x)$, depending on noncommuting coordinates,
can be mapped to a generalized product of these fields depending on usual coordinates by using the moyal
star product,

\begin{equation}
\left.\phi(\hat x)\psi(\hat x)=\phi(x) \ast \psi(x)=\exp\left(\frac{i}{2}\theta^{mn}
\frac{\partial}{\partial x^m}\frac{\partial}{\partial y^n}\right)\phi(x)\psi(y)\right|_{y \rightarrow x}
=\phi(x)\psi(x)+\frac{i}{2}\theta^{mn}\partial_m \phi(x) \partial_n \psi(x)+\mathcal{O}\left(\theta^2\right),
\label{moyal_product}
\end{equation}
which was originally considered in \cite{Moyal:1949sk}.
The extended expression appearing in ($\ref{moyal_product}$) of course depends on the noncommutativity
tensor $\theta$. Since the canonical formulation of general relativity is a field theory,
its setting on noncommutative space-time can also be represented on usual space-time by using the
moyal star product. This is considered in the next section. It is performed a calculation to the first
order in the noncommutativity parameter $\theta$.

\section{Canonical Gravity on Noncommutative Spacetime}

In the usual setting of the Hamiltonian formulation of general relativity, a foliation of space-time into
a time coordinate $\tau$ and a three-dimensional spacelike hypersurface $\Sigma$ has to be performed,
$\tau \times \Sigma$. This implies a splitting of the metric tensor $g_{\mu\nu}$ according to

\begin{equation}
g_{\mu\nu}=\left(\begin{matrix}N_a N^a-N^2 & N_b\\ N_c & h_{ab}\end{matrix}\right),
\label{splitted_metric}
\end{equation}
where $h_{ab}$ describes the three metric referring to the three-dimensional spacelike
hypersurface $\Sigma$, $N^a$ denotes the shift vector and $N$ describes the lapse function.
Based on this representation of the metric tensor and thus the gravitational field, all
other quantities are expressed.
The corresponding canonical gravity theory on noncommutative space-time is obtained by
replacing all usual quantities by quantities depending on noncommutative coordinates
and mapping the appearing products of these quantities to generalized products of
usual quantities by using the moyal star product. The moyal star product can be
represented as a series expansion in the noncommutativity parameter $\theta$ according to
the expression given in ($\ref{moyal_product}$). In this paper is performed a calculation to the first order in
the noncommutativity parameter $\theta$. In the considerations below, $A^{\theta 0}$ describes
the part of the generalization of an arbitrary quantity $A$, which does not depend on the
noncommutativity parameter $\theta$ and $A^{\theta 1}$ describes the part of the quantity
depending on the noncommutativity parameter $\theta$ to the first order.
All quantities appearing in the Hamiltonian formulation of general relativity corresponding
to the foliation of space-time and the corresponding splitting of the metric field
($\ref{splitted_metric}$) have to be transferred to noncommutative space-time and then
mapped to usual space-time. In this paper is made the presupposition that the
expressions obtained from the usual metric ($\ref{splitted_metric}$) are transferred
to noncommutative space and not already the products appearing in ($\ref{splitted_metric}$)
are transferred to noncommutative space-time. The generalized expressions on noncommutative
space-time are obtained by replacing the usual product between the quantities through the
star product ($\ref{moyal_product}$). If expressions already contain other quantities
containing products by themselves, then the corresponding generalized expressions have
to be inserted and of course all resulting terms of higher order have to be neglected.
Accordingly the Einstein-Hilbert action on noncommutative space-time reads

\begin{eqnarray}
S_{EH}^\theta&=&\frac{1}{16 \pi G}\int dt\ d^3 x \left(\pi^{ab}_\theta\ast \dot h_{ab}^{\theta}
-N\ast \bar {\mathcal{H}}_\mathcal{\tau}-N^a\ast \bar {\mathcal{H}}_a\right)
\nonumber\\
&=&\frac{1}{16 \pi G}\int dt\ d^3 x \left(\pi^{ab}_{\theta 0}\dot h_{ab}^{\theta 0}
+\pi^{ab}_{\theta 0}\dot h_{ab}^{\theta 1}
+\pi^{ab}_{\theta 1}\dot h_{ab}^{\theta 0}
+\frac{i}{2}\theta^{mn}\partial_m \pi^{ab}_{\theta 0} \partial_n h_{ab}^{\theta 0}
-N\bar {\mathcal{H}}_\mathcal{\tau}^{\theta 0}
-N\bar {\mathcal{H}}_\mathcal{\tau}^{\theta 1}
-\frac{i}{2}\theta^{mn}\partial_m N \partial_n \bar {\mathcal{H}}_\mathcal{\tau}^{\theta 0}
\right.\nonumber\\&&\left.\quad\quad\quad\quad\quad\quad\quad\quad
-N^a \bar {\mathcal{H}}_a^{\theta 0}
-N^a \bar {\mathcal{H}}_a^{\theta 1}
-\frac{i}{2}\theta^{mn}\partial_m N^a \partial_n \bar {\mathcal{H}}_a^{\theta 1}\right)
+\mathcal{O}\left(\theta^2\right),
\label{noncommutative_Einstein-Hilbert}
\end{eqnarray}
where $G$ denotes the gravitational constant.
Since the Einstein-Hilbert action in the canonical formulation on noncommutative space-time
($\ref{noncommutative_Einstein-Hilbert}$) already contains many quantities, which have to be generalized
on noncommutative space-time by themselves, $\pi^{ab}_\theta$, $\dot h_{ab}^\theta$, $\bar {\mathcal{H}}_\mathcal{\tau}^{\theta}$ and $\bar {\mathcal{H}}_a^{\theta}$ namely,
these generalized quantities have to be determined to make ($\ref{noncommutative_Einstein-Hilbert}$)
be defined explicitly. The generalized canonical conjugated variable to the three metric $h_{ab}$
appearing in ($\ref{noncommutative_Einstein-Hilbert}$) is given by the following expression:

\begin{eqnarray}
\pi^{ab}_{\theta}&=&\frac{\sqrt{h}}{16 \pi G}\ast \left(K^{ab}-K \ast h^{ab}\right)\nonumber\\
&=&\frac{\sqrt{h}}{16 \pi G}\left(K^{ab}-K h^{ab}\right)+\frac{i\theta^{mn}}{32\pi G}
\left(\frac{\partial_m h}{2\sqrt{h}}\partial_n K^{ab}-\frac{\partial_m h}{2\sqrt{h}}\partial_n K h^{ab}
-\frac{\partial_m h}{2\sqrt{h}} K \partial_n h^{ab}-\sqrt{h} \partial_m K \partial_n h^{ab}\right)
+\mathcal{O}\left(\theta^2\right)\nonumber\\
&\equiv&\pi^{ab}_{\theta 0}+\pi^{ab}_{\theta 1}+\mathcal{O}\left(\theta^2\right),
\end{eqnarray}
and the derivative of the three metric with respect to the chosen time coordinate
$\tau$ is given by

\begin{eqnarray}
\dot h_{ab}^{\theta}&=&32 \pi G N \ast \frac{1}{\sqrt{h}}\ast
\left(\pi_{ab}^{\theta}-\frac{1}{2} \pi^{cd}_\theta \ast h_{cd} \ast h_{ab}\right)+D_a^{\theta} \ast  N_b+D_b^{\theta} \ast N_a
\nonumber\\
&=&\frac{32 \pi G N}{\sqrt{h}}
\left[\left(\pi_{ab}^{\theta 0}+\pi_{ab}^{\theta 1}\right)
-\frac{1}{2}\left(\pi^{cd}_{\theta 0}+\pi^{cd}_{\theta 1}\right) h_{cd} h_{ab}\right]
+\partial_a N_b+\partial_b N_a
+\left(\omega_{aj}^{\theta 0\ i}+\omega_{aj}^{\theta 1\ i}\right)N_b
+\left(\omega_{bj}^{\theta 0\ i}+\omega_{bj}^{\theta 1\ i}\right)N_a\nonumber\\
&&+\frac{i}{2}\theta^{mn}\left(
-\partial_m N \frac{\partial_n h}{2h\sqrt{h}} \pi_{ab}^{\theta 0}
+\frac{\partial_m N}{\sqrt{h}} \partial_n \pi_{ab}^{\theta 0}
-N \frac{\partial_m h}{2h\sqrt{h}} \partial_n \pi_{ab}^{\theta 0}
-\partial_m N \frac{\partial_n h}{2h\sqrt{h}} \pi^{cd}_{\theta 0} h_{cd} h_{ab}
+\frac{\partial_m N}{\sqrt{h}} \partial_n \pi^{cd}_{\theta 0} h_{cd} h_{ab}
\right.\nonumber\\&&\left.
+\frac{\partial_m N}{\sqrt{h}} \pi^{cd}_{\theta 0} \partial_n h_{cd} h_{ab}
+\frac{\partial_m N}{\sqrt{h}} \pi^{cd}_{\theta 0} h_{cd} \partial_n h_{ab}
-N \frac{\partial_m h}{2h\sqrt{h}} \partial_n \pi^{cd}_{\theta 0} h_{cd} h_{ab}
-N \frac{\partial_n h}{2h\sqrt{h}} \pi^{cd}_{\theta 0} \partial_n h_{cd} h_{ab}
\right.\nonumber\\&&\left.
-N \frac{\partial_n h}{2h\sqrt{h}} \pi^{cd}_{\theta 0} h_{cd} \partial_n h_{ab}
+\frac{N}{\sqrt{h}} \partial_m \pi^{cd}_{\theta 0} \partial_n h_{cd} h_{ab}
+\frac{N}{\sqrt{h}} \partial_m \pi^{cd}_{\theta 0} h_{cd} \partial_n h_{ab}
+\frac{N}{\sqrt{h}} \pi^{cd}_{\theta 0} \partial_m h_{cd} \partial_n h_{ab}
\right.\nonumber\\&&\left.
+\partial_m \omega_{aj}^{\theta 0\ i}\partial_n N_b
+\partial_m \omega_{bj}^{\theta 0\ i}\partial_n N_a\right)
+\mathcal{O}\left(\theta^2\right)\nonumber\\
&\equiv&\dot h_{ab}^{\theta 0}+\dot h_{ab}^{\theta 1}+\mathcal{O}\left(\theta^2\right).
\label{dot_h}
\label{generalized_dot_h}
\end{eqnarray}
Since this expression ($\ref{generalized_dot_h}$) describing the generalized derivative of
the three metric $h_{ab}$ with respect to the time coordinate $\tau$, contains itself already
a generalized expression, the covariant derivative $D_a^{\theta}$ being defined by the generalized
spin connection $\omega_{aj}^{\theta\ i}$, this generalized quantity on noncommutative space-time has
to be calculated to make ($\ref{generalized_dot_h}$) be defined. The covariant derivative,
which has already been used in ($\ref{generalized_dot_h}$), applied to an arbitrary
vector is given by

\begin{eqnarray}
D_a^{\theta} \ast v^i&=&\partial_a v^i+\omega_{aj}^{i}\ast v^j
=\partial_a v^i+\omega_{aj}^{\theta 0\ i}v^j+\omega_{aj}^{\theta 1\ i}v^j
+\frac{i}{2}\theta^{mn}\partial_m \omega_{aj}^{\theta 0\ i}\partial_n v^j
+\mathcal{O}\left(\theta^2\right),
\label{covariant_derivative_noncommutative}
\end{eqnarray}
where the spin connection can be determined as

\begin{eqnarray}
\omega_{aj}^{\theta\ i}&=&e^d_k \ast e^f_j\ast e^i_c \ast \Gamma_{df}^c \ast e^k_a
-e^d_k \ast e^f_j \ast \partial_d e_f^i \ast e^k_a\nonumber\\
&=&e^d_k e^f_j e^i_c \Gamma_{df}^{\theta 0\ c} e^k_a
+e^d_k e^f_j e^i_c \Gamma_{df}^{\theta 1\ c} e^k_a
-e^d_k e^f_j \partial_d e_f^i e^k_a
\nonumber\\&&
+\frac{i}{2}\theta^{mn}\left(
\partial_m e^d_k \partial_n e^f_j e^i_c \Gamma_{df}^{\theta 0\ c} e^k_a
+\partial_m e^d_k e^f_j \partial_n e^i_c \Gamma_{df}^{\theta 0\ c} e^k_a
+\partial_m e^d_k e^f_j e^i_c \partial_n \Gamma_{df}^{\theta 0\ c} e^k_a
+\partial_m e^d_k e^f_j e^i_c \Gamma_{df}^{\theta 0\ c} \partial_n e^k_a
\right.\nonumber\\&&\left.
+e^d_k \partial_m e^f_j \partial_n e^i_c \Gamma_{df}^{\theta 0\ c} e^k_a
+e^d_k \partial_m e^f_j e^i_c \partial_n \Gamma_{df}^{\theta 0\ c} e^k_a
+e^d_k \partial_m e^f_j e^i_c \Gamma_{df}^{\theta 0\ c} \partial_n e^k_a
\right.\nonumber\\&&\left.
+e^d_k e^f_j \partial_m e^i_c \partial_n \Gamma_{df}^{\theta 0\ c} e^k_a
+e^d_k e^f_j \partial_m e^i_c \Gamma_{df}^{\theta 0\ c} \partial_n e^k_a
+e^d_k e^f_j e^i_c \partial_m \Gamma_{df}^{\theta 0\ c} \partial_n e^k_a
\right.\nonumber\\&&\left.
-\partial_m e^d_k \partial_n e^f_j \partial_d e_f^i e^k_a
-\partial_m e^d_k e^f_j \partial_n \partial_d e_f^i e^k_a
-\partial_m e^d_k e^f_j \partial_d e_f^i \partial_n e^k_a
\right.\nonumber\\&&\left.
-e^d_k \partial_m e^f_j \partial_n \partial_d e_f^i e^k_a
-e^d_k \partial_m e^f_j \partial_d e_f^i \partial_n e^k_a
-e^d_k e^f_j \partial_m \partial_d e_f^i \partial_n e^k_a\right)
+\mathcal{O}\left(\theta^2\right)\nonumber\\
&\equiv& \omega_{aj}^{\theta 0\ i}+\omega_{aj}^{\theta 1\ i}
+\mathcal{O}\left(\theta^2\right).
\label{generalized_spin_connection}
\end{eqnarray}
The generalized parts of the Christoffel symbols, $\Gamma_{ab}^{\theta 1\ c}$, appearing in 
($\ref{generalized_spin_connection}$), are defined below.
The parts of the Hamiltonian density on noncommutative space-time, $\bar {\mathcal{H}}_\mathcal{\tau}^\theta$
and $\bar {\mathcal{H}}_a^\theta$, are obtained in the same way by replacing the usual products through
star products,

\begin{eqnarray}
\bar {\mathcal{H}}_\mathcal{\tau}^{\theta}&=&16 \pi G\ G_{abcd}^{\theta}\ast \pi^{ab}_\theta\ast \pi^{cd}_\theta-\frac{\sqrt{h}}{16\pi G} \ast \left(R_\theta-2\Lambda\right),\nonumber\\
\bar {\mathcal{H}}_a^{\theta}&=&-2 D_b\ast p^{\theta\ b}_{\ \ a}.
\label{Hamiltonian_moyal}
\end{eqnarray}
If these star products are calculated explicitly, one obtains the following expressions:

\begin{eqnarray}
\bar {\mathcal{H}}_\mathcal{\tau}^{\theta}&=&16 \pi G\left[
G_{abcd}^{\theta 0}\left(\pi^{ab}_{\theta 0}\pi^{cd}_{\theta 0}
+2\pi^{ab}_{\theta 0}\pi^{cd}_{\theta 1}
\right)
+G_{abcd}^{\theta 1}\pi^{ab}_{\theta 0} \pi^{cd}_{\theta 0}
+\frac{i}{2}\theta^{mn}\left(
2\partial_m G_{abcd}^{\theta 0} \partial_n \pi^{ab}_{\theta 0} \pi^{cd}_{\theta 0}
+G_{abcd}^{\theta 0}\partial_m \pi^{ab}_{\theta 0} \partial_n \pi^{cd}_{\theta 0}\right)\right]
\nonumber\\&&\quad\quad\quad
-\frac{\sqrt{h}}{16\pi G}\left(R_{\theta 0}+R_{\theta 1}-2\Lambda\right)
-\frac{i\theta^{mn}}{32\pi G}\frac{\partial_m h}{2\sqrt{h}}\ \partial_n R_{\theta 0}+\mathcal{O}\left(\theta^2\right)\nonumber\\
&=&\bar {\mathcal{H}}_\mathcal{\tau}^{\theta 0}
+\bar {\mathcal{H}}_\mathcal{\tau}^{\theta 1}+\mathcal{O}\left(\theta^2\right),\nonumber\\
\bar {\mathcal{H}}_a^\theta&=&-2 \partial_b \pi^{\theta 0\ b}_{\ \ \ a}-2 \partial_b \pi^{\theta 1\ b}_{\ \ \ a}
-2 \omega_b^{\theta 0} \pi^{\theta 0\ b}_{\ \ \ a}-2 \omega_b^{\theta 1} \pi^{\theta 0\ b}_{\ \ \ a}
-2 \omega_b^{\theta 0} \pi^{\theta 1\ b}_{\ \ \ a}
-i\theta^{mn}\partial_m \omega_b^{\theta 0} \partial_n \pi^{\theta 0\ b}_{\ \ \ a}
+\mathcal{O}\left(\theta^2\right)\nonumber\\
&=&\bar {\mathcal{H}}_a^{\theta 0}+\bar {\mathcal{H}}_a^{\theta 1}+\mathcal{O}\left(\theta^2\right),
\label{hamiltonian_expressions}
\end{eqnarray}
where the generalized expression $G_{abcd}^\theta$ and its components read as following:

\begin{eqnarray}
G_{abcd}^\theta&=&\frac{1}{2\sqrt{h}}\ast \left(h^{ac}\ast h^{bd}+h^{ad}\ast h^{bc}-2h^{ab}\ast h^{cd}\right)\nonumber\\
&=&\frac{1}{2\sqrt{h}}\left(h^{ac} h^{bd}+h^{ad} h^{bc}-2h^{ab} h^{cd}\right)
+\frac{i\theta^{mn}}{4\sqrt{h}}\left(\partial_m h^{ac}\partial_n h^{bd}
+\partial_m h^{ad}\partial_n h^{bc}-2\partial_m h^{ab} \partial_n h^{cd}\right)\nonumber\\&&
-\frac{i\theta^{mn}}{8}\frac{\partial_m h}{h\sqrt{h}}\left(\partial_n h^{ac} h^{bd}+h^{ac} \partial_n h^{bd}
+\partial_n h^{ad} h^{bc}+h^{ad} \partial_n h^{bc}
-2 \partial_n h^{ab} h^{cd}-2 h^{ab} \partial_n h^{cd}\right)+\mathcal{O}\left(\theta^2\right)
\nonumber\\
&\equiv& G^{\theta 0}_{abcd}+G^{\theta 1}_{abcd}+\mathcal{O}\left(\theta^2\right).
\end{eqnarray}
In ($\ref{hamiltonian_expressions}$) has been used the symmetry $G_{abcd}^\theta=G_{cdab}^\theta$.
However, since in the Einstein-Hilbert action on noncommutative space-time ($\ref{noncommutative_Einstein-Hilbert}$),
the products between the Lagrange multipliers $N$ and $N^a$ and the components of the Hamiltonian density formulated
on noncommutative space-time, $\bar {\mathcal{H}}_\mathcal{\tau}^\theta$ and $\bar {\mathcal{H}}_a^\theta$, are
replaced by star products as well, leading to additional terms, the Hamiltonian and the diffeomorphism constraint
of the canonical formalism of gravity on noncommutative space-time are not determined completely by referring to them.
This is the reason why they have been written by using a bar. There have to be defined generalized expressions of the
components of the Hamiltonian density instead, which are obtained by varying the generalized Einstein-Hilbert action
with respect to the Lagrange multipliers $N$ and $N^a$. This leads to the complete Hamiltonian and diffeomorphism
constraint in the noncommutative canonical gravity theory,

\begin{eqnarray}
\mathcal{H}_{\mathcal{\tau}}^\theta&=&-\frac{\delta S_{EH}^\theta}{\delta N}
=\bar {\mathcal{H}}_\mathcal{\tau}^{\theta 0}+\bar {\mathcal{H}}_\mathcal{\tau}^{\theta 1}
-\frac{i}{2}\theta^{mn}\partial_m \partial_n \bar {\mathcal{H}}_\mathcal{\tau}^{\theta 0}
+\mathcal{O}\left(\theta^2\right)=0,\nonumber\\
\mathcal{H}_a^\theta&=&-\frac{\delta S_{EH}^\theta}{\delta N^a}
=\bar {\mathcal{H}}_a^{\theta 0}+\bar {\mathcal{H}}_a^{\theta 1}
-\frac{i}{2}\theta^{mn} \partial_m \partial_n \bar {\mathcal{H}}_a^{\theta 0}
+\mathcal{O}\left(\theta^2\right)=0.
\label{constraints_hamiltonian_gravity}
\end{eqnarray}
To obtain the generalized expression of the Ricci-scalar appearing in the components of the Hamiltonian,
the Riemann tensor on noncommutative space-time has to be determined and therefore the corresponding
Christoffel symbols, which have already been used in ($\ref{generalized_spin_connection}$), have to
be calculated first. They are given by

\begin{eqnarray}
\Gamma_{ab}^{c\ \theta}&=&
\frac{1}{2} h^{cd} \ast \left(\partial_a h_{db}+\partial_b h_{da}-\partial_d h_{ab}\right)\nonumber\\
&=&\frac{1}{2} h^{cd} \left(\partial_a h_{db}+\partial_b h_{da}-\partial_d h_{ab}\right)
+\frac{i}{4}\theta^{mn} \partial_m h^{cd} \left(\partial_n \partial_a h_{db}
+\partial_n \partial_b h_{da}-\partial_n \partial_d h_{ab}\right)+\mathcal{O}\left(\theta^2\right)
\nonumber\\
&\equiv&\Gamma_{\ \ ab}^{\theta 0\ c}+\Gamma_{\ \ ab}^{\theta 1\ c}+\mathcal{O}\left(\theta^2\right).
\end{eqnarray}
The Riemann tensor accordingly reads as follows:

\begin{eqnarray}
R_{abc}^{\theta\ \ d}&=&\partial_a \Gamma_{bc}^{\theta\ d}-\partial_b \Gamma_{ac}^{\theta\ d}
+\Gamma_{ac}^{\theta\ e}\ast \Gamma_{be}^{\theta\ d}
-\Gamma_{bc}^{\theta\ e}\ast \Gamma_{ae}^{\theta\ d}\nonumber\\
&=&\partial_a \Gamma_{\ \ bc}^{\theta 0\ d}
+\partial_a \Gamma_{\ \ bc}^{\theta 1\ d}
-\partial_b \Gamma_{\ \ ac}^{\theta 0\ d}
-\partial_b \Gamma_{\ \ ac}^{\theta 1\ d}\nonumber\\&&
+\Gamma_{\ \ ac}^{\theta 0\ e}\Gamma_{\ \ be}^{\theta 0\ d}
+\Gamma_{\ \ ac}^{\theta 0\ e}\Gamma_{\ \ be}^{\theta 1\ d}
+\Gamma_{\ \ ac}^{\theta 1\ e}\Gamma_{\ \ be}^{\theta 0\ d}
-\Gamma_{\ \ bc}^{\theta 0\ e}\Gamma_{\ \ ae}^{\theta 0\ d}
-\Gamma_{\ \ bc}^{\theta 0\ e}\Gamma_{\ \ ae}^{\theta 1\ d}
-\Gamma_{\ \ bc}^{\theta 1\ e}\Gamma_{\ \ ae}^{\theta 0\ d}
\nonumber\\
&&+\frac{i}{2}\theta^{mn}\left[\partial_m \Gamma_{\ \ ac}^{\theta 0\ e}\partial_n \Gamma_{\ \ be}^{\theta 0\ d}
-\partial_m \Gamma_{\ \ bc}^{\theta 0\ e}\partial_n\Gamma_{\ \ ae}^{\theta 0\ d}\right]
+\mathcal{O}\left(\theta^2\right),
\end{eqnarray}
and this means that the Ricci tensor reads as follows:

\begin{eqnarray}
R_{ab}^{\theta}&=&\partial_a \Gamma_{cb}^{\theta\ c}-\partial_c \Gamma_{ab}^{\theta\ c}
+\Gamma_{ab}^{\theta\ e}\ast \Gamma_{ce}^{\theta\ c}
-\Gamma_{cb}^{\theta\ e}\ast \Gamma_{ae}^{\theta\ c}\nonumber\\
&=&\partial_a \Gamma_{\ \ cb}^{\theta 0\ c}
+\partial_a \Gamma_{\ \ cb}^{\theta 1\ c}
-\partial_c \Gamma_{\ \ ab}^{\theta 0\ c}
-\partial_c \Gamma_{\ \ ab}^{\theta 1\ c}
\nonumber\\&&
+\Gamma_{\ \ ab}^{\theta 0\ e}\Gamma_{\ \ ce}^{\theta 0\ c}
+\Gamma_{\ \ ab}^{\theta 0\ e}\Gamma_{\ \ ce}^{\theta 1\ c}
+\Gamma_{\ \ ab}^{\theta 1\ e}\Gamma_{\ \ ce}^{\theta 0\ c}
-\Gamma_{\ \ cb}^{\theta 0\ e}\Gamma_{\ \ ae}^{\theta 0\ c}
-\Gamma_{\ \ cb}^{\theta 0\ e}\Gamma_{\ \ ae}^{\theta 1\ c}
-\Gamma_{\ \ cb}^{\theta 1\ e}\Gamma_{\ \ ae}^{\theta 0\ c}
\nonumber\\
&&+\frac{i}{2}\theta^{mn}\left[\partial_m\Gamma_{\ \ ab}^{\theta 0\ e}\partial_n \Gamma_{\ \ ce}^{\theta 0\ c}
-\partial_m\Gamma_{\ \ cb}^{\theta 0\ e}\partial_n\Gamma_{\ \ ae}^{\theta 0\ c}\right]
+\mathcal{O}\left(\theta^2\right).
\end{eqnarray}
The Ricci scalar can finally be determined to

\begin{eqnarray}
R_{\theta}&=&h^{ab}\ast R_{ab}^{\theta}=h^{ab}\ast\left[\partial_a \Gamma_{cb}^{\theta\ c}-\partial_c \Gamma_{ab}^{\theta\ c}
+\Gamma_{ab}^{\theta\ e}\ast \Gamma_{ce}^{\theta\ c}
-\Gamma_{cb}^{\theta\ e}\ast \Gamma_{ae}^{\theta\ c}\right]\nonumber\\
&=&h^{ab}\left[\partial_a \Gamma_{\ \ cb}^{\theta 0\ c}
+\partial_a \Gamma_{\ \ cb}^{\theta 1\ c}
-\partial_c \Gamma_{\ \ ab}^{\theta 0\ c}
-\partial_c \Gamma_{\ \ ab}^{\theta 1\ c}\right.\nonumber\\&&\left.
+\Gamma_{\ \ ab}^{\theta 0\ e} \Gamma_{\ \ ce}^{\theta 0\ c}
+\Gamma_{\ \ ab}^{\theta 0\ e} \Gamma_{\ \ ce}^{\theta 1\ c}
+\Gamma_{\ \ ab}^{\theta 1\ e} \Gamma_{\ \ ce}^{\theta 0\ c}
-\Gamma_{\ \ cb}^{\theta 0\ e} \Gamma_{\ \ ae}^{\theta 0\ c}
-\Gamma_{\ \ cb}^{\theta 0\ e} \Gamma_{\ \ ae}^{\theta 1\ c}
-\Gamma_{\ \ cb}^{\theta 1\ e} \Gamma_{\ \ ae}^{\theta 0\ c}
\right]\nonumber\\
&&+\frac{i}{2}\theta^{mn}h^{ab}\left[\partial_m\Gamma_{\ \ ab}^{\theta 0\ e}\partial_n\Gamma_{\ \ ce}^{\theta 0\ c}
-\partial_m\Gamma_{\ \ cb}^{\theta 0\ e}\partial_n\Gamma_{\ \ ae}^{\theta 0\ c}\right]
\nonumber\\&&
+\frac{i}{2}\theta^{mn}\partial_m h_{ab}
\left[\partial_n \partial_a \Gamma_{\ \ cb}^{\theta 0\ c}
-\partial_n \partial_c \Gamma_{\ \ ab}^{\theta 0\ c}
+\partial_n \Gamma_{\ \ ab}^{\theta 0\ e} \Gamma_{\ \ ce}^{\theta 0\ c}
-\partial_n \Gamma_{\ \ cb}^{\theta 0\ e} \Gamma_{\ \ ae}^{\theta 0\ c}\right.\nonumber\\
&&\left.+\Gamma_{\ \ ab}^{\theta 0\ e} \partial_n \Gamma_{\ \ ce}^{\theta 0\ c}
-\Gamma_{\ \ cb}^{\theta 0\ e} \partial_n \Gamma_{\ \ ae}^{\theta 0\ c}\right]
+\mathcal{O}\left(\theta^2\right)\nonumber\\
&\equiv&R_{\theta 0}+R_{\theta 1}+\mathcal{O}\left(\theta^2\right).
\end{eqnarray}
This means that the decisive quantities of the canonical formulation of canonical gravity on noncommutative
space-time have been represented on commutative space-time by using the moyal star product ($\ref{moyal_product}$)
and presupposing a consideration to the first order in $\theta$. To obtain the corresponding quantum theory,
it is inevitable first to consider the general idea of canonical quantization on noncommutative space-time,
which is directly related to the generalization of the formulation of commutators on noncommutative space-time.

\section{Canonical Quantization on Noncommutative Space-Time}

To obtain the corresponding quantum theory of a theory on noncommutative space-time by referring to canonical quantization,
the quantization rules on a noncommutative space-time have to be determined first. The appearance of a generalization of
the quantization of a theory arises from the generalized properties of commutators according to the star product.
The transition to noncommuting coordinates implies an analogous transition of the commutator between to arbitrary
field operators $\hat A$ and $\hat B$:

\begin{equation}
\left[\hat A(x), \hat B(y)\right] \longrightarrow \left[\hat A(\hat x),\hat B(\hat y)\right],
\label{transition_commutator}
\end{equation}
where $\left[\hat A,\hat B\right]=\hat A \hat B-\hat B \hat A$.
The generalized commutator appearing in ($\ref{transition_commutator}$) can be calculated in dependence on
the usual commutator and the noncommutativity parameter $\theta$ by using the moyal product
($\ref{moyal_product}$) as follows:

\begin{eqnarray}
\left[\hat A(\hat x),\hat B(\hat y)\right]&=&
\hat A(\hat x)\hat B(\hat y)-\hat B(\hat y)\hat A(\hat x)
=\left[\hat A(x),\hat B(y)\right]_{\ast}
=\hat A(x) \ast \hat B(y)-\hat B(y) \ast \hat A(x)\nonumber\\
&&=\hat A(x) \hat B(y)-\hat B(y) \hat A(x)+\left[\frac{i}{2}\theta^{mn}\partial_m \hat A(x) \partial_n \hat B(y)
-\frac{i}{2}\theta^{mn}\partial_m \hat B(y) \partial_n \hat A(x)\right]\delta(x-y)
+\mathcal{O}\left(\theta^2\right)
\nonumber\\
&&=\hat A(x) \hat B(y)-\hat B(y) \hat A(x)+\left[\frac{i}{2}\theta^{mn}\partial_m \hat A(x) \partial_n \hat B(y) +\frac{i}{2}\theta^{mn}\partial_n \hat B(y) \partial_m \hat A(x)\right]\delta(x-y)
+\mathcal{O}\left(\theta^2\right)
\nonumber\\
&&=\left[\hat A(x),\hat B(y)\right]
+\left[\frac{i}{2}\theta^{mn}\left\{\partial_m \hat A(x),\partial_n \hat B(x)\right\}\right]\delta(x-y)
+\mathcal{O}\left(\theta^2\right),
\label{generalized_commutator}
\end{eqnarray}
where $\left\{\hat A,\hat B\right\}=\hat A \hat B+\hat B \hat A$. This means for the canonical quantization
of a scalar field $\phi$ and its canonical conjugated quantity $\pi_\phi$,

\begin{equation}
\left[\hat \phi(x),\hat \pi_\phi(y)\right]=i\delta(x-y) \longrightarrow 
\left[\hat \phi(x),\hat \pi_\phi(y)\right]=\left[1+\frac{1}{2}\theta^{mn}
\left\{\partial_m \hat \phi(x),\partial_n \hat \pi_\phi(x)\right\}\right]i\delta(x-y).
\label{quantization_scalar}
\end{equation}
This generalization of the quantization rule according to ($\ref{quantization_scalar}$) can be realized by the following
representation of the operators $\hat \phi$ and $\hat \pi_\phi$:

\begin{eqnarray}
\hat \phi(x)|\Psi[\phi]\rangle=\phi(x)|\Psi[\phi]\rangle,\quad \hat \pi_\phi(x)|\Psi[\phi]\rangle
&=&-i\left[\frac{\delta}{\delta \phi(x)}+\frac{i}{2}\theta^{mn}\partial_m \phi(x)\int d^3 z\ \partial_n \delta(z-x)
\frac{\delta^2}{\delta [\phi(z)]^2}\right]|\Psi[\phi]\rangle
+\mathcal{O}\left(\theta^2\right).\nonumber\\
\label{representation_scalar}
\end{eqnarray}
The validity of the representation of the canonical conjugated field operator in ($\ref{representation_scalar}$)
can be seen, if one calculates the commutator between the field operator and its canonical conjugated operator
in the representation ($\ref{representation_scalar}$),

\begin{eqnarray}
&&\left[\hat \phi(x),\hat \pi_\phi(y)\right]=
\left[\phi(x),-i\left\{\frac{\delta}{\delta \phi(y)}+\frac{i}{2}\theta^{mn}\partial_m \phi(y)\int d^3 z\ \partial_n \delta(z-y)
\frac{\delta^2}{\delta [\phi(z)]^2}\right\}\right]+\mathcal{O}\left(\theta^2\right)\nonumber\\&&
=i\left[\delta(x-y)+i\theta^{mn}\partial_m \phi(y)\int d^3 z\ \partial_n \delta(z-y)
\delta(x-z)\frac{\delta}{\delta [\phi(z)]}\right]+\mathcal{O}\left(\theta^2\right)\nonumber\\&&
=i\left[\delta(x-y)+i\theta^{mn}\partial_m \phi(y)\partial_n \delta(x-y)\frac{\delta}{\delta [\phi(x)]}\right]
+\mathcal{O}\left(\theta^2\right)\nonumber\\&&
=i\delta(x-y)\left[1-i\theta^{mn}\partial_m \phi(x)\partial_n\left(\frac{\delta}{\delta [\phi(x)]}\right)\right]
+\mathcal{O}\left(\theta^2\right),
\label{calculation_commutator}
\end{eqnarray}
where has been used the relation $\partial_m \delta(x-y)f(x)=-\delta(x-y)\partial_m f(x)$.
If one now inserts the representation of the operators ($\ref{representation_scalar}$) to the right
hand side of ($\ref{quantization_scalar}$), one obtains:

\begin{eqnarray}
&&\left[1+\frac{1}{2}\theta^{mn}\left\{\partial_m \hat \phi(x),\partial_n \hat \pi(x)\right\}\right]i\delta(x-y)
=i\delta(x-y)\left[1-\frac{i}{2}\theta^{mn}\left\{\partial_m \phi(x),\partial_n\left(\frac{\delta}{\delta [\phi(x)]}\right)\right\}\right]+\mathcal{O}\left(\theta^2\right)\nonumber\\
&&=i\delta(x-y)\left[1-i\theta^{mn}\partial_m \phi(x)\partial_n\left(\frac{\delta}{\delta [\phi(x)]}\right)\right]+\mathcal{O}\left(\theta^2\right).
\label{inserting_representation}
\end{eqnarray}
Thus both sides of ($\ref{quantization_scalar}$) are equal and therefore the commutator is fulfilled by the
representation ($\ref{representation_scalar}$). In the second step of ($\ref{inserting_representation}$),
the anti-commutator has been replaced by the doubled product of the expressions within it. 
The possibility to perform this replacement arises from the vanishing variation of the derivative of a
field with respect to the field what can be seen as follows:

\begin{eqnarray}
\frac{\delta \partial_m \phi(x)}{\delta \phi(x)}
&=&\frac{\delta}{\delta \phi(x)}\int d^3 y\ \delta(x-y)\partial_m \phi(y)
=-\frac{\delta}{\delta \phi(x)}\int d^3 y\ \partial_m \delta(x-y) \phi(y)\nonumber\\
&=&-\int d^3 y\ \partial_m \delta(x-y) \frac{\delta \phi(y)}{\delta \phi(x)}
=-\int d^3 y\ \partial_m \delta(x-y) \delta(x-y)
=-\partial_m \delta(0)=0.
\end{eqnarray}
Because of the generalization of the expression of an arbitrary commutator according to the noncommutativity
of the coordinates ($\ref{generalized_commutator}$), the commutation relations between the Hamiltonian
expressions have to be generalized as well. If the Hamiltonian of an arbitrary field theory is given by

\begin{equation}
\hat H_\theta=\int d^3 x \left(N \ast \hat {\bar {\mathcal{H}}}_{\tau}^{\theta}
+N^a \ast \hat {\bar {\mathcal{H}}}_a^{\theta}\right),
\end{equation}
then the commutation relations between $\hat {\bar {\mathcal{H}}}_{\tau}^\theta$ and
$\hat {\bar {\mathcal{H}}}_a^\theta$ on noncommutative space-time are given by

\begin{eqnarray}
\left[\hat{\bar {\mathcal{H}}}_{\tau}^\theta(x),\hat{\bar {\mathcal{H}}}_{\tau}^\theta(y)\right]&=&\partial_a \delta(x-y)
\left(h^{ab}(x)\hat {\bar {\mathcal{H}}}_b^\theta(x)+h^{ab}(y)\hat{\bar {\mathcal{H}}}_b^\theta(y)\right)
+\left[\frac{i}{2}\theta^{mn}\left\{\partial_m \hat{\bar {\mathcal{H}}}_{\tau}^\theta(x),
\partial_n \hat{\bar {\mathcal{H}}}_{\tau}^\theta(y)\right\}\right]\delta(x-y)+\mathcal{O}\left(\theta^2\right)
,\nonumber\\
\left[\hat{\bar {\mathcal{H}}}_{a}^\theta(x),\hat{\bar {\mathcal{H}}}_{\tau}^\theta(y)\right]&=&
\hat {\mathcal{H}}_{\tau}^\theta\partial_a \delta(x-y)
+\left[\frac{i}{2}\theta^{mn}\left\{\partial_m \hat{\bar {\mathcal{H}}}_{a}^\theta(x),
\partial_n \hat{\bar {\mathcal{H}}}_{\tau}^\theta(y)\right\}\right]\delta(x-y)+\mathcal{O}\left(\theta^2\right)
,\nonumber\\
\left[\hat{\bar {\mathcal{H}}}_{a}^\theta(x),\hat{\bar {\mathcal{H}}}_{b}^\theta(y)\right]&=&
\hat{\bar {\mathcal{H}}}_b^\theta(x)\partial_a \delta(x-y)
+\hat{\bar {\mathcal{H}}}_a^\theta(y)\partial_b \delta(x-y)
+\left[\frac{i}{2}\theta^{mn}\left\{\partial_m \hat{\bar {\mathcal{H}}}_a^\theta(x),
\partial_n \hat{\bar {\mathcal{H}}}_b^\theta(y)\right\}\right]\delta(x-y)+\mathcal{O}\left(\theta^2\right).
\end{eqnarray}
Remember that the expressions $\hat {\bar {\mathcal{H}}}_{\tau}^\theta$ as well as $\hat {\bar {\mathcal{H}}}_a^\theta$
determine the Hamiltonian and the diffeomorphism constraint with respect to the relations given in ($\ref{constraints_hamiltonian_gravity}$).

\section{Quantum Geometrodynamics on Noncommutative Space-Time}

If the dynamical quantities of general relativity on noncommutative space-time shall be quantized canonically,
the usual commutator has to be generalized according to the general transition rule, ($\ref{transition_commutator}$)
with ($\ref{generalized_commutator}$). This leads to the following quantization condition for the variables
of quantum geometrodynamics, $h_{ab}$ and $\pi^{ab}$:

\begin{equation}
\left[\hat h_{ab}(x),\hat \pi^{cd}(y)\right]=i\left[\delta^c_{(a} \delta^d_{b)}
+\frac{1}{2}\theta^{mn}\left\{\partial_m \hat h_{ab}(x),\partial_n \hat \pi^{cd}(y)\right\}
\right]\delta(x-y)
+\mathcal{O}\left(\theta^2\right).
\label{quantization_gravity}
\end{equation}
The corresponding representation of the operators appearing in ($\ref{quantization_gravity}$) in the three
metric space is given by

\begin{eqnarray}
\hat h_{ab}(x)|\Psi[h] \rangle&=&h_{ab}(x)|\Psi[h] \rangle,\quad\nonumber\\
\hat \pi^{ab}(x)|\Psi[h] \rangle&=&-i\left[\frac{\delta}{\delta h_{ab}(x)}
+\frac{i}{2}\theta^{mn}\mathcal{X}^{ijab}_{efgh}\partial_m h_{ij}(x)\int d^3 z\ \partial_n \delta\left(z-x\right)
\frac{\delta}{\delta h_{ef}(z)}\frac{\delta}{\delta h_{gh}(z)}\right]
|\Psi[h] \rangle+\mathcal{O}\left(\theta^2\right),
\label{operators_gravity}
\end{eqnarray}
where has been introduced the new tensor $\mathcal{X}$, which is a tensor of eight order over the vector
space of three vectors referring to the three dimensional submanifold $\Sigma$. If $A$ and $B$ are assumed
to be two arbitrary tensors of second order over the same vector space, then the components of
$\mathcal{X}$, $\mathcal{X}^{abcd}_{efgh}$, are defined by the following relations:

\begin{eqnarray}
\mathcal{X}^{abcd}_{efgh} A_{ab} B^{ef}&\equiv& A_{gh} B^{cd},\quad
\mathcal{X}^{abcd}_{efgh} A_{ab} B^{ef}=\mathcal{X}^{abcd}_{ghef} A_{ab} B^{ef}.
\label{definition_X}
\end{eqnarray}
The validity of the representation of the operators of quantum geometrodynamics on noncommutative space-time,
which is given in ($\ref{operators_gravity}$), is shown in analogy to ($\ref{calculation_commutator}$)
and ($\ref{inserting_representation}$) by calculating the commutator between $\hat h_{ab}(x)$ and
$\hat \pi^{cd}(y)$ in the representation ($\ref{operators_gravity}$),

\begin{eqnarray}
\left[\hat h_{ab}(x),\hat \pi^{cd}(y)\right]&=&\left[h_{ab}(x),-i\left\{\frac{\delta}{\delta h_{cd}(y)}
+\frac{i}{2}\theta^{mn}\mathcal{X}^{ijcd}_{efgh}\partial_m h_{ij}(y)\int d^3 z\ \partial_n \delta\left(z-y\right)
\frac{\delta}{\delta h_{ef}(z)}\frac{\delta}{\delta h_{gh}(z)}\right\}\right]+\mathcal{O}\left(\theta^2\right)\nonumber\\
&=&i\left\{\frac{\delta h_{ab}(x)}{\delta h_{cd}(y)}
+i\theta^{mn}\mathcal{X}^{ijcd}_{efgh}\partial_m h_{ij}(y)\int d^3 z\ \partial_n \delta\left(z-y\right)
\left(\frac{\delta h_{ab}(x)}{\delta h_{ef}(z)}\frac{\delta}{\delta h_{gh}(z)}\right)
\right\}+\mathcal{O}\left(\theta^2\right)\nonumber\\
&=&i\left\{\delta_{(a}^c \delta_{b)}^d \delta(x-y)
+i\theta^{mn}\mathcal{X}^{ijcd}_{efgh}\partial_m h_{ij}(y)\int d^3 z\ \partial_n \delta\left(z-y\right)
\left(\delta_{(a}^e \delta_{b)}^f \delta(x-z)\frac{\delta}{\delta h_{gh}(z)}\right)
\right\}+\mathcal{O}\left(\theta^2\right)\nonumber\\
&=&i\left\{\delta_{(a}^c \delta_{b)}^d \delta(x-y)
+i\theta^{mn}\mathcal{X}^{ijcd}_{efgh}\partial_m h_{ij}(y)\partial_n \delta\left(x-y\right)
\left(\delta_{(a}^e \delta_{b)}^f \frac{\delta}{\delta h_{gh}(x)}\right)\right\}+\mathcal{O}\left(\theta^2\right)\nonumber\\
&=&i\left\{\delta_{(a}^c \delta_{b)}^d \delta(x-y)
+i\theta^{mn}\mathcal{X}^{ijcd}_{abgh}\partial_m h_{ij}(y)\partial_n \delta\left(x-y\right)
\left(\frac{\delta}{\delta h_{gh}(x)}\right)\right\}+\mathcal{O}\left(\theta^2\right)\nonumber\\
&=&i\left\{\delta_{(a}^c \delta_{b)}^d \delta(x-y)
+i\theta^{mn}\partial_m h_{ab}(y)\partial_n \delta\left(x-y\right)
\left(\frac{\delta}{\delta h_{cd}(x)}\right)\right\}+\mathcal{O}\left(\theta^2\right)\nonumber\\
&=&i\delta(x-y)\left\{\delta_{(a}^c \delta_{b)}^d-i\theta^{mn}\partial_m h_{ab}(x)\partial_n\left(\frac{\delta}{\delta h_{cd}(x)}\right)\right\}+\mathcal{O}\left(\theta^2\right),
\label{calculation_commutator_quantum_geometrodynamics}
\end{eqnarray}
where again has been used the relation $\partial_m \delta(x-y)f(x)=-\delta(x-y)\partial_m f(x)$, and inserting ($\ref{operators_gravity}$) also to the right hand side of ($\ref{quantization_gravity}$),

\begin{eqnarray}
&&\left[\delta_{(a}^c \delta_{b)}^d
+\frac{1}{2}\theta^{mn}\left\{\partial_m \hat h_{ab}(x),\partial_n \hat \pi^{cd}(x)\right\}\right]i\delta(x-y)
=i\delta(x-y)\left[\delta_{(a}^c \delta_{b)}^d-\frac{i}{2}\theta^{mn}\left\{\partial_m h_{ab}(x),\partial_n\left(\frac{\delta}{\delta h_{cd}(x)}
\right)\right\}\right]+\mathcal{O}\left(\theta^2\right),\nonumber\\
&&=i\delta(x-y)\left[\delta_{(a}^c \delta_{b)}^d
-i\theta^{mn}\partial_m h_{ab}(x)\partial_n\left(\frac{\delta}{\delta h_{cd}(x)}\right)\right]+\mathcal{O}\left(\theta^2\right).
\label{inserting_representation_quantum_geometrodynamics}
\end{eqnarray}
Since the result of ($\ref{calculation_commutator_quantum_geometrodynamics}$) is equal to the result of ($\ref{inserting_representation_quantum_geometrodynamics}$), ($\ref{operators_gravity}$) yields the correct
representation of the operators defined by ($\ref{quantization_gravity}$).
To obtain the subspace of states, which are dynamically valid, the constraints have to be implemented,
which are obtained by converting the classical constraints ($\ref{constraints_hamiltonian_gravity}$) to
operators, which act on the quantum states referring to the gravitational field. And to perform this procedure,
the Hamiltonian expressions reformulated on noncommutative space-time ($\ref{hamiltonian_expressions}$)
have to be inserted to the constraints on noncommutative space-time ($\ref{constraints_hamiltonian_gravity}$).
Remember that the generalized Hamiltonian expressions are not equivalent with the generalized constraints,
because of the star product with the Lagrange multipliers within the generalized action. After this the
transition takes place by replacing the dynamical quantities $h_{ab}$ and $\pi^{ab}$ within the classical
constraints, ($\ref{constraints_hamiltonian_gravity}$) with ($\ref{hamiltonian_expressions}$) inserted,
by the corresponding operators and accordingly leads to the following quantum constraints:

\begin{eqnarray}
\hat {\mathcal{H}}_\mathcal{\tau}^\theta|\Psi[h] \rangle&=&\left\{
16 \pi G\left[\hat G_{abcd}^{\theta 0}\left(\hat \pi^{ab}_{\theta 0}\hat \pi^{cd}_{\theta 0}
+2\hat \pi^{ab}_{\theta 0}\hat \pi^{cd}_{\theta 1}\right)
+\hat G_{abcd}^{\theta 1}\hat \pi^{ab}_{\theta 0} \hat \pi^{cd}_{\theta 0}
+\frac{i}{2}\theta^{mn}\left(
2\partial_m \hat G_{abcd}^{\theta 0} \partial_n \hat \pi^{ab}_{\theta 0} \hat \pi^{cd}_{\theta 0}
+\hat G_{abcd}^{\theta 0}\partial_m \hat \pi^{ab}_{\theta 0} \partial_n \hat \pi^{cd}_{\theta 0}\right)\right]
\right.\nonumber\\&&\left.
-\frac{\sqrt{\hat h}}{16\pi G}\left(\hat R_{\theta 0}+\hat R_{\theta 1}-2\Lambda\right)
-\frac{i\theta^{mn}}{32\pi G}\frac{\partial_m \hat h}{2\sqrt{\hat h}}\partial_n \hat R_{\theta 0}
\right.\nonumber\\&&\left.
-\frac{i}{2}\theta^{mn}\partial_m \partial_n \left[
16 \pi G\ \hat G_{abcd}^{\theta 0} \hat \pi^{ab}_{\theta 0} \hat \pi^{cd}_{\theta 0}
-\frac{\sqrt{\hat h}}{16\pi G} \left(\hat R_{\theta 0}-2\Lambda\right)\right]
\right\}|\Psi[h] \rangle+\mathcal{O}\left(\theta^2\right)=0,
\nonumber\\
\hat {\mathcal{H}}_a^\theta|\Psi[h] \rangle&=&
\left[-2 \partial_b \hat \pi^{\theta 0\ b}_{\ \ \ a}-2 \partial_b \hat \pi^{\theta 1\ b}_{\ \ \ a}
-2 \hat \omega_b^{\theta 0} \hat \pi^{\theta 0\ b}_{\ \ \ a}-2 \hat \omega_b^{\theta 1} \hat \pi^{\theta 0\ b}_{\ \ \ a}
-2 \hat \omega_b^{\theta 0} \hat \pi^{\theta 1\ b}_{\ \ \ a}
-i\theta^{mn}\partial_m \hat \omega_b^{\theta 0} \partial_n \hat \pi^{\theta 0\ b}_{\ \ \ a}
\right.\nonumber\\&&\left.
-\frac{i}{2}\theta^{mn}\partial_m \partial_n \left(
-2 \partial_b \hat \pi^{\theta 0\ b}_{\ \ \ a}
-2 \hat \omega_b^{\theta 0} \hat \pi^{\theta 0\ b}_{\ \ \ a}
\right)\right]
|\Psi[h] \rangle+\mathcal{O}\left(\theta^2\right)=0.
\label{constraints_gravitational_field}
\end{eqnarray}
If the concrete representations of the operators given in ($\ref{operators_gravity}$) are inserted to
($\ref{constraints_gravitational_field}$), then the quantum constraints of canonical quantum gravity
on noncommutative space-time to the first order in the noncommutativity parameter $\theta$
read as follows:

\begin{eqnarray}
\hat {\mathcal{H}}_\mathcal{\tau}^\theta(x)|\Psi[h] \rangle&=&\left\{
16 \pi G\left[G_{abcd}^{\theta 0}(x)\left(-\frac{\delta}{\delta h_{cd}(x)}
-\left[i\theta^{mn}\mathcal{X}^{ijcd}_{efgh}\partial_m h_{ij}(x)\int d^3 z\ \partial_n \delta\left(z-x\right)
\frac{\delta}{\delta h_{ef}(z)}\frac{\delta}{\delta h_{gh}(z)}\right]
\right)\frac{\delta}{\delta h_{ab}(x)}
\right.\right.\nonumber\\&&\left.\left.
-\frac{i}{2}\theta^{mn}\partial_m G_{abcd}^{\theta 0}(x)\left(
2\partial_n \frac{\delta}{\delta h_{ab}(x)}\frac{\delta}{\delta h_{cd}(x)}
+\partial_m \frac{\delta}{\delta h_{ab}(x)} \partial_n \frac{\delta}{\delta h_{cd}(x)}
\right)
-G_{abcd}^{\theta 1}(x)\frac{\delta}{\delta h_{ab}(x)}\frac{\delta}{\delta h_{cd}(x)}\right]
\right.\nonumber\\&&\left.
-\frac{\sqrt{h(x)}}{16\pi G}\left(R_{\theta 0}(x)+R_{\theta 1}(x)-2\Lambda\right)
-\frac{i\theta^{mn}}{32\pi G}\frac{\partial_m h(x)}{2\sqrt{h(x)}}\partial_n R_{\theta 0}(x)
\right.\nonumber\\&&\left.
-\frac{i}{2}\theta^{mn}\partial_m \partial_n \left[
-16 \pi G\ G_{abcd}^{\theta 0}(x) \frac{\delta}{\delta h_{ab}(x)}\frac{\delta}{\delta h_{cd}(x)}
-\frac{\sqrt{h(x)}}{16\pi G} \left(R_{\theta 0}(x)-2\Lambda\right)\right]
\right\}
|\Psi[h] \rangle+\mathcal{O}\left(\theta^2\right)=0,
\nonumber\\
\hat {\mathcal{H}}_a^\theta(x)|\Psi[h] \rangle&=&
i\left\{2 \partial_b \frac{\delta}{\delta h^a_b(x)}
+2 \partial_b \left[\frac{i}{2}\theta^{mn}\mathcal{X}^{ijab}_{efgh}\partial_m h_{ij}(x)\int d^3 z\ \partial_n \delta\left(z-x\right)\frac{\delta}{\delta h_{ef}(z)}\frac{\delta}{\delta h_{gh}(z)}\right]
\right.\nonumber\\&&\left.
+2 \omega_b^{\theta 0}(x)\left[\frac{i}{2}\theta^{mn}\mathcal{X}^{ijab}_{efgh}\partial_m h_{ij}(x)\int d^3 z\ \partial_n \delta\left(z-x\right)\frac{\delta}{\delta h_{ef}(z)}\frac{\delta}{\delta h_{gh}(z)}\right]
+\left[2 \omega_b^{\theta 0}(x)+2 \omega_b^{\theta 1}(x)\right]\frac{\delta}{\delta h^a_b(x)}
\right.\nonumber\\&&\left.
+i\theta^{mn}\partial_m \omega_b^{\theta 0}(x) \partial_n \frac{\delta}{\delta h^a_b(x)}
-\frac{i}{2}\theta^{mn}\partial_m \partial_n \left(
2 \partial_b \frac{\delta}{\delta h_b^a(x)} 
+2 \omega_b^{\theta 0} \frac{\delta}{\delta h_b^a(x)}
\right)
\right\}
|\Psi[h] \rangle
+\mathcal{O}\left(\theta^2\right)=0.
\label{quantum_constraints_gravitational_field}
\end{eqnarray}
Note, that the dependence of the field quantities on the coordinate is explicitly written in all expressions 
like ($\ref{quantum_constraints_gravitational_field}$) containing the representation of the operators of the scalar
field or the gravitational field respectively. This is because the representation of the canonical conjugated quantity to
the fields contains an integral over the variation with respect to the field on a certain point and therefore
the variations with respect to the fields depend on the parameter the integral refers to instead of the point
the complete field operator representation refers to. In case of all the other expressions all quantities
directly depend on the same space-time point parameter and therefore it has not to be written explicitly.

\section{Coupling to Matter}

In the last section quantum geometrodynamics on noncommutative space-time has been formulated.
But the coupling of matter fields to the gravitational field has been omitted so far. Therefore
the question remains how the canonical quantum theory of gravity on noncommutative space-time
looks like under incorporation of the coupling to matter fields. In this section an arbitrary
scalar field coupled to the gravitational field shall be considered. To incorporate the
quantum description of a scalar field to the quantum description of general relativity according
to quantum geometrodynamics, it has to be treated in the Hamiltonian formulation as well.
In the canonical formulation a scalar field on noncommutative space-time is described by
the following Hamiltonian:

\begin{eqnarray}
H_\phi^\theta&=&\int d^3 x\left[N\ast \left(\frac{1}{2\sqrt{h}}\ast \pi_\phi+\frac{\sqrt{h}}{2}\ast h^{ab}
\ast \partial_a \phi \ast \partial_b \phi
+\frac{m^2}{2} \sqrt{h} \ast \phi^2 \right)+ N^a\ast \pi_\phi \ast \partial_a \phi\right].\nonumber\\
&=&\int d^3 x\left[N \left[\frac{\pi_\phi^{\theta 0}+\pi_\phi^{\theta 1}}{2\sqrt{h}}+\frac{\sqrt{h}}{2} h^{ab}
\partial_a \phi \partial_b \phi+\frac{m^2}{2} \sqrt{h} \phi^2\right]
+N^a \left(\pi_\phi^{\theta 0}+\pi_\phi^{\theta 1}\right) \partial_a \phi
\right.\nonumber\\&&\left.
+\frac{i}{4}\theta^{mn}\left(-\partial_m N \frac{\partial_n h}{2h\sqrt{h}} \pi_\phi^{\theta 0}
+\frac{\partial_m N}{\sqrt{h}} \partial_n \pi_\phi^{\theta 0}
-N \frac{\partial_m h}{2h\sqrt{h}} \partial_n \pi_\phi^{\theta 0}
+\partial_m N \frac{\partial_n h}{2\sqrt{h}} h^{ab} \partial_a \phi \partial_b \phi
+\partial_m N \sqrt{h} \partial_n h^{ab} \partial_a \phi \partial_b \phi
\right.\right.\nonumber\\&&\left.\left.
+2\partial_m N \sqrt{h} h^{ab} \partial_n \partial_a \phi \partial_b \phi
+N \frac{\partial_m h}{2\sqrt{h}}\partial_n h^{ab} \partial_a \phi \partial_b \phi
+2N \frac{\partial_m h}{2\sqrt{h}} h^{ab} \partial_n \partial_a \phi \partial_b \phi
+2N \sqrt{h} \partial_m h^{ab} \partial_n \partial_a \phi \partial_b \phi
\right.\right.\nonumber\\&&\left.\left.
+N \sqrt{h} h^{ab} \partial_m \partial_a \phi \partial_n \partial_b \phi
+m^2 \partial_m N \frac{\partial_n h}{2\sqrt{h}} \phi^2
+m^2 \partial_m N \sqrt{h} \partial_n \phi^2
+m^2 N \frac{\partial_m h}{2\sqrt{h}} \partial_n \phi^2
\right.\right.\nonumber\\&&\left.\left.
+2\partial_m N^a \partial_n \pi_\phi^{\theta 0} \partial_a \phi
+2\partial_m N^a \pi_\phi^{\theta 0} \partial_n \partial_a \phi
+2N^a \partial_m \pi_\phi^{\theta 0} \partial_n \partial_a \phi\right)\right]
+\mathcal{O}\left(\theta^2\right).
\label{complete_Hamiltonian_matter_field}
\end{eqnarray}
The corresponding constraints are obtained by varying the Hamiltonian describing a scalar field
on noncommutative space-time ($\ref{complete_Hamiltonian_matter_field}$) with respect to the
Lagrange multipliers. This is completely analogue to the case referring to the gravitational field
($\ref{constraints_hamiltonian_gravity}$) with the difference that here the components 
of the usual Hamiltonian density on noncommutative space-time, which correspond to
($\ref{hamiltonian_expressions}$), have not been determined separately. Thus the constraints
in its classical manifestation read 

\begin{equation}
\mathcal{H}_{\tau\phi}^\theta=\frac{\delta H_\phi^\theta}{\delta N}=0,\quad \mathcal{H}_{a\phi}^\theta
=\frac{\delta H_\phi^\theta}{\delta N^a}=0,
\end{equation}
and if the variations with respect to the Lagrange multipliers are calculated explicitly,
this leads to the following classical constraints for the scalar field:

\begin{eqnarray}
\mathcal{H}_{\tau\phi}^\theta&=&\frac{\left(\pi_\phi^{\theta 0}+\pi_\phi^{\theta 1}\right)}{2\sqrt{h}}
+\frac{\sqrt{h}}{2} h^{ab}\partial_a \phi \partial_b \phi+\frac{1}{2}m^2 \sqrt{h} \phi^2
\nonumber\\&&
+\frac{i}{4}\theta^{mn}\left[-\partial_m\left(-\frac{\partial_n h}{2h\sqrt{h}} \pi_\phi^{\theta 0}\right)
-\partial_m \left(\frac{1}{\sqrt{h}} \partial_n \pi_\phi^{\theta 0}\right)
-\frac{\partial_m h}{2h\sqrt{h}} \partial_n \pi_\phi^{\theta 0}
-\partial_m\left(\frac{\partial_n h}{2\sqrt{h}} h^{ab} \partial_a \phi \partial_b \phi\right)
-\partial_m\left(\sqrt{h} \partial_n h^{ab} \partial_a \phi \partial_b \phi\right)
\right.\nonumber\\&&\left.\quad\quad\quad\quad
-2\partial_m\left(\sqrt{h} h^{ab} \partial_n \partial_a \phi \partial_b \phi\right)
+\frac{\partial_m h}{2\sqrt{h}} \partial_n h^{ab} \partial_a \phi \partial_b \phi
+2\frac{\partial_m h}{2\sqrt{h}} h^{ab} \partial_n \partial_a \phi \partial_b \phi
+2\sqrt{h} \partial_m h^{ab} \partial_n \partial_a \phi \partial_b \phi
\right.\nonumber\\&&\left.\quad\quad\quad\quad
+\sqrt{h} h^{ab} \partial_m \partial_a \phi \partial_n \partial_b \phi
-m^2 \partial_m \left(\partial_n \sqrt{h} \phi^2\right)
-m^2 \partial_m \left(\sqrt{h} \partial_n \phi^2\right)
+m^2 \frac{\partial_m h}{2\sqrt{h}}\ \partial_n \phi^2
\right]+\mathcal{O}\left(\theta^2\right)=0,
\nonumber\\
\mathcal{H}_{a\phi}^\theta&=&\left(\pi_\phi^{\theta 0}+\pi_\phi^{\theta 1}\right) \partial_a \phi
+\frac{i}{2}\theta^{mn}\left[
-\partial_m\left(\partial_n \pi_\phi^{\theta 0} \partial_a \phi\right)
-\partial_m\left(\pi_\phi^{\theta 0} \partial_n \partial_a \phi\right)
+\partial_m \pi_\phi^{\theta 0} \partial_n \partial_a \phi\right]+\mathcal{O}\left(\theta^2\right)=0.
\label{constraints_scalar_field}
\end{eqnarray}
These constraints can in complete analogy to the case of the gravitational field,
($\ref{constraints_gravitational_field}$) and ($\ref{quantum_constraints_gravitational_field}$),
be converted to the corresponding quantum constraints by using the quantization rule
($\ref{quantization_scalar}$) and the corresponding representation of the operators
($\ref{representation_scalar}$) and inserting them into ($\ref{constraints_scalar_field}$).
This leads to the following expressions for the quantum constraints referring to
the scalar field:

\begin{eqnarray}
&&\hat {\mathcal{H}}_\phi^\theta(x)|\Psi[h,\phi]\rangle=\left\{\frac{-i}{2\sqrt{h}}
\left[\frac{\delta}{\delta \phi(x)}+\frac{i}{2}\theta^{mn}\partial_m \phi(x)\int d^3 z\ \partial_n \delta(z-x)
\frac{\delta^2}{\delta [\phi(z)]^2}\right]
+\frac{\sqrt{h}}{2} h^{ab}(x)\partial_a \phi(x) \partial_b \phi(x)+\frac{1}{2}m^2 \sqrt{h(x)} [\phi(x)]^2
\right.\nonumber\\&&\left.
+\frac{i}{4}\theta^{mn}\left[i\partial_m\left(\partial_n \frac{1}{\sqrt{h(x)}}\frac{\delta}{\delta \phi(x)}\right)
+i\partial_m \left(\frac{1}{\sqrt{h(x)}} \partial_n \frac{\delta}{\delta \phi(x)}\right)
+i\frac{\partial_m h(x)}{2h\sqrt{h(x)}} \partial_n \frac{\delta}{\delta \phi(x)}
-\partial_m\left(\partial_n \sqrt{h(x)} h^{ab}(x) \partial_a \phi(x) \partial_b \phi(x)\right)
\right.\right.\nonumber\\&&\left.\left.
-\partial_m\left(\sqrt{h(x)} \partial_n h^{ab}(x) \partial_a \phi(x) \partial_b \phi(x)\right)
-2\partial_m\left(\sqrt{h(x)} h^{ab}(x) \partial_n \partial_a \phi(x) \partial_b \phi(x)\right)
+\frac{\partial_m h(x)}{2\sqrt{h(x)}} \partial_n h^{ab}(x) \partial_a \phi(x) \partial_b \phi(x)
\right.\right.\nonumber\\&&\left.\left.
+2\frac{\partial_m h(x)}{2\sqrt{h(x)}} h^{ab}(x) \partial_n \partial_a \phi(x) \partial_b \phi(x)
+2\sqrt{h(x)} \partial_m h^{ab} \partial_n \partial_a \phi(x) \partial_b \phi(x)
+\sqrt{h(x)} h^{ab}(x) \partial_m \partial_a \phi(x) \partial_n \partial_b \phi(x)
\right.\right.\nonumber\\&&\left.\left.
-m^2 \partial_m \left(\frac{\partial_n h(x)}{2\sqrt{h(x)}} [\phi(x)]^2\right)
-m^2 \partial_m \left(\sqrt{h(x)} \partial_n [\phi(x)]^2\right)
+m^2 \frac{\partial_m h(x)}{2\sqrt{h(x)}}\partial_n [\phi(x)]^2
\right]\right\}|\Psi[h,\phi]\rangle+\mathcal{O}\left(\theta^2\right)=0,
\nonumber\\
&&\hat {\mathcal{H}}_{a\phi}^\theta(x)|\Psi[h,\phi]\rangle=
\left\{-i\left[\frac{\delta}{\delta \phi(x)}+
\frac{i}{2}\theta^{mn}\partial_m \phi(x)\int d^3 z\ \partial_n \delta(z-x)
\frac{\delta^2}{\delta [\phi(z)]^2}
\right]\partial_a \phi(x)
\right.\nonumber\\&&\left.
+\frac{i}{2}\theta^{mn}\left[
\partial_m\left(i\partial_n \frac{\delta}{\delta \phi(x)}\partial_a \phi(x)\right)
+\partial_m\left(i\frac{\delta}{\delta \phi(x)} \partial_n \partial_a \phi(x)\right)
-i\partial_m \frac{\delta}{\delta \phi(x)} \partial_n \partial_a \phi(x)\right]\right\}
|\Psi[h,\phi]\rangle+\mathcal{O}\left(\theta^2\right)=0.
\label{quantum_constraints_matter_field}
\end{eqnarray}
The complete constraints are then obtained by adding the quantum constraints of the matter field
($\ref{quantum_constraints_matter_field}$) to the quantum constraints of the gravitational field  
($\ref{quantum_constraints_gravitational_field}$) and leads to the following equations:

\begin{equation}
\left[\hat {\mathcal{H}}_\tau^\theta(x)+\hat {\mathcal{H}}_{\phi\tau}^\theta(x)\right]|\Psi[h,\phi]\rangle=0,\quad
\left[\hat {\mathcal{H}}_{a}^\theta(x)+\hat {\mathcal{H}}_{\phi a}^\theta(x)\right]|\Psi[h,\phi]\rangle=0,
\end{equation}
where the components of the Hamiltonian density are defined explicitly in ($\ref{quantum_constraints_gravitational_field}$)
and ($\ref{quantum_constraints_matter_field}$). This means that for the case of a scalar field the complete
constraints containing the interaction between the scalar fields as matter fields and the gravitational field
have been obtained. This holds at least for the framework of quantum geometrodynamics. In the next section the
description of canonical quantum gravity on noncommutative space-time will be extended to the
Ashtekar formalism on which loop quantum gravity is based on.

\section{Transition to the Ashtekar Formalism}

The description of canonical quantum gravity on noncommutative space-time can analogously be obtained within
the setting using Ashtekars variables. In this formalism a special representation of the spin connection is the
decisive dynamical quantity, which is defined by the relation

\begin{equation}
A_a^i=\frac{\Gamma_a^i+\beta K_a^i}{G},
\end{equation}
where $\beta$ denotes the Immirzi parameter and $\Gamma_a^i$ is defined by the relation

\begin{equation}
\Gamma_a^i=-\frac{1}{2}\omega_{ajk}\epsilon^{ijk}.
\end{equation}
$\epsilon^{ijk}$ denotes as usual the complete anti-symmetric tensor of third order. The canonical conjugated
quantity to the connection is given by a generalized version of the triad. If the triad is defined by the
relation  

\begin{equation}
h^{ab}=\delta^{ij}e_i^a e_j^b,
\end{equation}
then the generalized triad representing the canonical conjugated quantity to the connection is given by

\begin{equation}
E_i^a=\sqrt{h}e_i^a.
\end{equation}
To obtain the generalized quantum theory on noncommutative space-time, the quantization principle
has to be generalized in analogy to quantum geometrodynamics ($\ref{quantization_gravity}$) and
thus in accordance with the generalization of commutators on noncommutative space-time given in
($\ref{generalized_commutator}$). This means that the commutation relation between the operator
referring to the connection, $\hat A_a^i$, and the operator referring to the canonical conjugated
quantity, $\hat E_i^a$, has to be assumed to be of the following shape:

\begin{equation}
\left[\hat A_a^i(x),\hat E_j^b(y)\right]=i\left[8\pi\beta \delta^i_j \delta^b_a+\frac{1}{2}\theta^{mn}
\left\{\partial_m \hat A_a^i,\partial_n \hat E_j^b\right\}\right]
\delta(x-y)+\mathcal{O}\left(\theta^2\right).
\label{quantization_Ashtekar}
\end{equation}
The corresponding representation of the operators fulfilling the generalized quantization principle
($\ref{quantization_Ashtekar}$) in the connection space is in analogy to ($\ref{operators_gravity}$)
given by

\begin{eqnarray}
\hat A_a^i(x)|\Psi[A] \rangle&=&A_a^i(x)|\Psi[A] \rangle,\nonumber\\
\hat E_i^a(x)|\Psi[A] \rangle&=&-i\left[8\pi\beta\frac{\delta}{\delta A_a^i(x)}
+\frac{i}{2}\theta^{mn}\delta^{kp}\delta^{lq}\delta_{wz}\delta_{ir}\mathcal{X}^{ezar}_{cpdq}\partial_m A_e^w(x)
\right.\nonumber\\&&\left.\quad\quad\quad\quad\quad\quad\quad\quad
\times \int d^3 z\ \partial_n \delta(z-x)\frac{\delta}{\delta A_c^k(z)}\frac{\delta}{\delta A_d^l(z)}
\right]|\Psi[A] \rangle
+\mathcal{O}\left(\theta^2\right)=0.
\label{operators_Ashtekar}
\end{eqnarray}
The calculation of the validity of the representation ($\ref{operators_Ashtekar}$) is completely isomorphic
to the case of quantum geometrodynamics given in ($\ref{calculation_commutator_quantum_geometrodynamics}$)
and ($\ref{inserting_representation_quantum_geometrodynamics}$) and therefore it is not written down separately.
To obtain the generalized quantum Hamiltonian and diffeomorphism constraint, the field strength tensor referring
to the connection $A_a^i$ on noncommutative space-time has to be determined first. By using again the moyal
star product ($\ref{moyal_product}$) the generalized field strength tensor can be calculated to

\begin{eqnarray}
F_{ab}^{\theta\ i}&=&2G\partial_a A_b^i-2G\partial_b A_a^i+G^2 \epsilon^i_{\ jk} A_a^j \ast A_b^k\nonumber\\
&=&2G\partial_a A_b^i-2G\partial_b A_a^i+G^2 \epsilon^i_{\ jk} A_a^j A_b^k
+\frac{i}{2}\theta^{mn} G^2 \epsilon^i_{\ jk}\partial_m A_a^j \partial_n A_b^k+\mathcal{O}\left(\theta^2\right)
\equiv F_{\ \ ab}^{\theta 0\ i}+F_{\ \ ab}^{\theta 1\ i}+\mathcal{O}\left(\theta^2\right).
\end{eqnarray}
In the Ashtekar formalism appears besides the Hamiltonian and the diffeomorphism constraint, the Gauss constraint,
which has to be generalized as well. In the usual setting it is given by

\begin{equation}
\mathcal{G}=\mathcal{D}_a E^a_i=\partial_a E^a_i+G \epsilon_{ijk} A^j_a E^{ka}=0.
\label{usual_Gauss_constraint}
\end{equation}
In case of noncommutative geometry the Gauss constraint ($\ref{usual_Gauss_constraint}$) has to be generalized to

\begin{eqnarray}
\mathcal{G}_\theta &\equiv& \mathcal{D}_a \ast E^a_i=\partial_a E^a_i+G\epsilon_{ijk}A^j_a\ast E^{ka}
=\partial_a E^a_i+G\epsilon_{ijk}A^j_a E^{ka}
+\frac{i}{2}\theta^{mn}G\epsilon_{ijk}\partial_m A^j_a \partial_n E^{ka}+\mathcal{O}\left(\theta^2\right)=0,
\nonumber\\
\Leftrightarrow\quad \mathcal{G}&=&\mathcal{D}_a E^a_i=\partial_a E^a_i+G\epsilon_{ijk}A^j_a E^{ka}=
-\frac{i}{2}\theta^{mn}G\epsilon_{ijk}\partial_m A^j_a \partial_n E^{ka}+\mathcal{O}\left(\theta^2\right).
\label{generalized_Gauss_constraint}
\end{eqnarray}
The quantum theoretical version of the Gauss constraint in case of a noncommutative space-time
($\ref{generalized_Gauss_constraint}$) is accordingly obtained by converting the variables of
gravity in the Ashtekar formulation appearing in ($\ref{generalized_Gauss_constraint}$) to the
corresponding operators and inserting ($\ref{operators_Ashtekar}$) to ($\ref{generalized_Gauss_constraint}$).
This procedure leads to the following manifestation of the generalized quantum Gauss constraint:

\begin{eqnarray}
&&\hat {\mathcal{G}}(x)|\Psi[A]\rangle \equiv \hat {\mathcal{D}}_a \hat E^a_i(x)|\Psi[A]\rangle\nonumber\\
&&=\left\{-i{\mathcal{D}}_a \left[8\pi\beta\frac{\delta}{\delta A_a^i(x)}
+\frac{i}{2}\theta^{mn}\delta^{kp}\delta^{lq}\delta_{wz}\delta_{ir}\mathcal{X}^{ezar}_{cpdq}\partial_m A_e^w(x)\int d^3 z\ \partial_n \delta(z-x)\frac{\delta}{\delta A_c^k(z)}\frac{\delta}{\delta A_d^l(z)}
\right]
\right\}|\Psi[A]\rangle+\mathcal{O}\left(\theta^2\right)\nonumber\\
&&=-\left[\frac{8\pi\beta}{2}\theta^{mn}G\epsilon_{ijk}\partial_m A^j_a(x) \partial_n \frac{\delta}{\delta A_{ka}(x)}\right]
|\Psi[A]\rangle+\mathcal{O}\left(\theta^2\right).
\end{eqnarray}
The components of the Hamiltonian in the Ashtekar formalism formulated on noncommutative space-time read

\begin{eqnarray}
\bar {\mathcal{H}}_\theta&=&\epsilon^{ijk}F_{abk}^{\theta}\ast E_i^a \ast E_j^b
=\epsilon^{ijk}F_{abk}^{\theta 0} E_i^a E_j^b
+\epsilon^{ijk}F_{abk}^{\theta 1} E_i^a E_j^b
+i\theta^{mn}\epsilon^{ijk}\partial_m F_{abk}^{\theta 0}\partial_n E_i^a E_j^b
\nonumber\\&&\quad\quad\quad\quad\quad\quad\quad\quad\quad\quad\quad
+\frac{i}{2}\theta^{mn}\epsilon^{ijk}F_{abk}^{\theta 0}\partial_m E_i^a \partial_n E_j^b
+\mathcal{O}\left(\theta^2\right)
=\bar {\mathcal{H}}^{\theta 0}+\bar {\mathcal{H}}^{\theta 1}+\mathcal{O}\left(\theta^2\right),
\nonumber\\
\bar {\mathcal{H}}_a^\theta&=&F_{ab}^{\theta\ i}\ast E^b_i
=F_{ab}^{\theta 0\ i} E^b_i+F_{ab}^{\theta 1\ i} E^b_i
+\frac{i}{2}\theta^{mn} \partial_m F_{ab}^{\theta 0\ i} \partial_n E^b_i+\mathcal{O}\left(\theta^2\right)
=\bar {\mathcal{H}}_a^{\theta 0}+\bar {\mathcal{H}}_a^{\theta 1}+\mathcal{O}\left(\theta^2\right).
\label{barHamiltonian}
\end{eqnarray}
But as in case of quantum geometrodynamics the components of the Hamiltonian formulated on noncommutative
space-time do not represent the complete constraints on noncommutative space-time, since because of the
moyal product with the Lagrange multipliers, the complete constraints are given by
($\ref{constraints_hamiltonian_gravity}$). Accordingly the complete constraints can be determined
by inserting ($\ref{barHamiltonian}$) to ($\ref{constraints_hamiltonian_gravity}$) and accordingly
they read as follows:

\begin{eqnarray}
{\mathcal{H}}_\theta
&=&\left[\epsilon^{ijk} F_{abk}^{\theta 0} E_i^a E_j^b
+\epsilon^{ijk} F_{abk}^{\theta 1} E_i^a E_j^b\right.
+i\theta^{mn}\epsilon^{ijk}\partial_m F_{abk}^{\theta 0}\partial_n E_i^a E_j^b
\nonumber\\&&\left.
+\frac{i}{2}\theta^{mn}\epsilon^{ijk} F_{abk}^{\theta 0}\partial_m E_i^a \partial_n E_j^b
-\frac{i}{2}\theta^{mn}\partial_m \partial_n
\left(\epsilon^{ijk} F_{abk}^{\theta 0} E_i^a E_j^b\right)
\right]+\mathcal{O}\left(\theta^2\right)=0,
\nonumber\\
{\mathcal{H}}_a^\theta
&=&\left[F_{ab}^{\theta 0\ i} E^b_i+F_{ab}^{\theta 1\ i} E^b_i
+\frac{i}{2}\theta^{mn} \partial_m F_{ab}^{\theta 0\ i} \partial_n E^b_i
-\frac{i}{2}\theta^{mn}\partial_m \partial_n
\left(F_{ab}^{\theta 0\ i} E^b_i\right)\right]
+\mathcal{O}\left(\theta^2\right)=0.
\end{eqnarray}
The corresponding quantum constraints are then obtained by converting the dynamical quantities to operators
and using ($\ref{operators_Ashtekar}$) in the resulting expression,

\begin{eqnarray}
\hat {\mathcal{H}}_\theta(x)|\Psi[A]\rangle
&=&-\left\{\epsilon^{ijk} F_{abk}^{\theta 0}(x) \frac{\delta}{\delta A_a^i(x)}\frac{\delta}{\delta A_b^j(x)}
+\epsilon^{ijk} F_{abk}^{\theta 1}(x) \frac{\delta}{\delta A_a^i(x)} \frac{\delta}{\delta A_b^j(x)}
+i\theta^{mn}\epsilon^{ijk} \partial_m F_{abk}^{\theta 0}(x)
\partial_n \frac{\delta}{\delta A_a^i(x)} \frac{\delta}{\delta A_b^j(x)}
\right.\\&&\left.
+\epsilon^{ijk} F_{abk}^{\theta 0}(x)
\left[\theta^{mn}\delta^{kp}\delta^{lq}\delta_{wz}\delta_{ir}\mathcal{X}^{ezar}_{cpdq}\partial_m A_e^w(x)\int d^3 z\ \partial_n \delta(z-x)\frac{\delta}{\delta A_c^k(z)}\frac{\delta}{\delta A_d^l(z)}\right]
\frac{\delta}{\delta A_b^j(x)}
\right.\nonumber\\&&\left.
+\frac{i}{2}\theta^{mn}\epsilon^{ijk} F_{abk}^{\theta 0}(x)\partial_m \frac{\delta}{\delta A_a^i(x)}
\partial_n \frac{\delta}{\delta A_b^j(x)}
-\frac{i}{2}\theta^{mn}\partial_m \partial_n
\left(\epsilon^{ijk} F_{abk}^{\theta 0} \frac{\delta}{\delta A_a^i(x)}
\frac{\delta}{\delta A_b^j(x)}\right)
\right\}|\Psi[A]\rangle+\mathcal{O}\left(\theta^2\right)=0,
\nonumber\\
\hat {\mathcal{H}}_a^\theta(x)|\Psi[A]\rangle
&=&-i\left\{F_{ab}^{\theta 0\ i}(x) \frac{\delta}{\delta A_b^i(x)}
-iF_{ab}^{\theta 0\ i}
\left[\frac{i}{2}\theta^{mn}\delta^{kp}\delta^{lq}\delta_{wz}\delta_{ir}\mathcal{X}^{ezar}_{cpdq}\partial_m A_e^w(x)\int d^3 z\ \partial_n \delta(z-x)\frac{\delta}{\delta A_c^k(z)}\frac{\delta}{\delta A_d^l(z)}\right]
\right.\nonumber\\&&\left.
+F_{ab}^{\theta 1\ i} \frac{\delta}{\delta A_b^i(x)}
+\frac{i}{2}\theta^{mn} \partial_m F_{ab}^{\theta 0\ i}(x) \partial_n \frac{\delta}{\delta A_b^i(x)}
-\frac{i}{2}\theta^{mn}\partial_m \partial_n
\left(F_{ab}^{\theta 0\ i}(x) \frac{\delta}{\delta A_b^i(x)}\right)\right\}|\Psi[A]\rangle
+\mathcal{O}\left(\theta^2\right)=0.\nonumber
\end{eqnarray}

\section{Holonomy Representation and Area Operator}

In the last section canonical quantum gravity based on Ashtekars variables on noncommutative space-time
has been explored. In this section is considered the holonomy representation of the gravitational field
corresponding to the Ashtekar formalism as it is used in loop quantum gravity. This means that the generalized
quantum description of general relativity developed in the last sections is considered with respect to the holonomy representation. A generalization of a quantum description of general relativity in the context of loop quantum
gravity being based on a generalization of the description of the classical metrical structure has been explored
in \cite{Ma:2011aa}. However, by referring to this holonomy representation and presupposing the generalized operators
of Ashtekars variables on noncommutative space-time, which have been given in ($\ref{operators_Ashtekar}$)
and represent the generalized quantization rule ($\ref{quantization_Ashtekar}$), it is possible to
calculate a generalized area operator according to the generalization of the quantum description of
general relativity induced by the noncommutativity of space-time.
The area operator and the volume operator in the context of loop quantum gravity have been considered
in \cite{Rovelli:1994ge} for the first time. Further studies concerning the area operator can be found in \cite{Ashtekar:1996eg},\cite{AmelinoCamelia:1998hk},\cite{Engle:2007mu} and concerning the
volume operator in \cite{Brunnemann:2004xi},\cite{Meissner:2005mx},\cite{Brunnemann:2007ca},\cite{Brunneman:2007as},\cite{Brunemann:2011xd}.
In this paper the considerations will remain restricted to the area operator. The holonomy corresponding to
the connection $A_a^i$ representing the gravitational field along a curve described by the coordinates
$\gamma^a(s)$ in dependence on the parameter $s$ is given by

\begin{equation}
U\left[A,\gamma\right]\left(0,\lambda\right)=\mathcal{P}\exp\left(G\int_0^\lambda ds
\frac{d \gamma^a(s)}{ds}A_a^i(\gamma(s))\tau_i\right),
\label{holonomy}
\end{equation}
where the $\tau_i$ describe the generators of the $SU(2)$. Accordingly for the holonomy holds:
$U\left[A,\gamma\right](s) \in SU(2)$, $U\left[A,\gamma\right](0)=\bf{1}$ and it fulfils
the following relation:

\begin{equation}
\frac{d}{ds}U\left[A,\gamma\right](s)-G\frac{d\gamma^a(s)}{ds} A_a^i(\gamma(s))\tau_i U\left[A,\gamma\right](s)=0.
\end{equation}
The smeared out version of the canonical conjugated operator to the connection operator, $\hat E^a_i$ namely,
where it is integrated over a surface $\mathcal{S}$ embedded in the space-like three-dimensional submanifold
$\Sigma$, is in its generalized shape on noncommutative space-time being based on the representation
($\ref{operators_Ashtekar}$) given by the following expression:

\begin{eqnarray}
\hat E_i(\mathcal{S})&=&-i \int_{\mathcal{S}}d \sigma^1 d \sigma^2\
n_a(\vec \sigma)\left[8\pi\beta \frac{\delta}{\delta A_a^i[x(\vec \sigma)]}
+\frac{i}{2}\theta^{mn}\delta^{kp}\delta^{lq}\delta_{wz}\delta_{ir}\mathcal{X}^{ezar}_{cpdq}
\partial_m A_e^w[x(\vec \sigma)]
\right.\nonumber\\&&\left.\quad\quad\quad\quad\quad\quad\quad\quad\quad\quad\quad\quad\quad\quad\quad\quad\quad
\times \int d^3 z\ \partial_n \delta(z-x(\vec \sigma))\frac{\delta}{\delta A_c^k(z)}
\frac{\delta}{\delta A_d^l(z)}\right],
\label{smeared_area}
\end{eqnarray}
where the three vector $n_a(\vec \sigma)$ is defined as follows:

\begin{equation}
n_a(\vec \sigma)=\epsilon_{abc}\frac{\partial x^b(\vec \sigma)}{\partial \sigma_1}
\frac{\partial x^c(\vec \sigma)}{\partial \sigma_2}.
\end{equation}
The embedding is defined by the mapping:
$\left(\sigma_1,\sigma_2\right) \rightarrow x^a\left(\sigma_1,\sigma_2\right)=x^a\left(\vec \sigma\right)$.
To obtain the expression, which arises, if the smeared out version of the canonical conjugated operator is
applied to the holonomy of the connection and thus of the gravitational field, it is necessary to consider
the application of the derivative with respect to the connection to the holonomy,
which is determined by the relation

\begin{equation}
\frac{\delta U[A,\gamma]}{\delta A_a^i(x)}=G \int_\gamma
ds \frac{d \gamma^a(s)}{ds}\delta\left(x-\gamma(s)\right)
U\left[A,\gamma_1\right]\tau_i U\left[A,\gamma_2\right],
\label{relation_dUdA}
\end{equation}
where $\gamma_1$ and $\gamma_2$ denote the two sections of the curve $\gamma$, if it is divided by the point,
where $x$ is equal to $\gamma(s)$. The relation ($\ref{relation_dUdA}$) has been calculated in 
\cite{Lewandowski:1993zq} and can also be found in \cite{Rovelli:2004} and \cite{Kiefer:2004}.
It shall now be considered the application of ($\ref{smeared_area}$) to the holonomy along a curve $\gamma$ under the
precondition that $\gamma$ intersects $\mathcal{S}$ at one single point $p$. By using the relation
($\ref{relation_dUdA}$), it is possible to calculate the expression arising through application of the operator
($\ref{smeared_area}$) to the holonomy ($\ref{holonomy}$). Since in ($\ref{smeared_area}$) the derivation with
respect to the connection appears quadratically, the calculation becomes much more intricated in case of
a noncommutative space-time as it is presupposed here. Application of the operator ($\ref{smeared_area}$)
to the holonomy ($\ref{holonomy}$) yields the following expression:

\begin{eqnarray}
\hat E_i(\mathcal{S})U\left[A,\alpha\right]&=&-i \int_{\mathcal{S}}d \sigma^1 d \sigma^2\
\epsilon_{abc}\frac{\partial x^a(\vec \sigma)}{\partial \sigma_1}
\frac{\partial x^b(\vec \sigma)}{\partial \sigma_2}
\left\{8\pi\beta \frac{\delta U\left[A,\alpha\right]}{\delta A_c^i[x(\vec \sigma)]}
\right.\nonumber\\&&\left.
+i\theta^{mn}\delta^{kp}\delta^{lq}\delta_{wz}\delta_{ir}\mathcal{X}^{ezcr}_{gphq}\partial_m A_e^w[x(\vec \sigma)]\int d^3 z\ \partial_n \delta\left(z-x(\vec \sigma)\right)\frac{\delta}{\delta A_g^k[z]}\frac{\delta U\left[A,\alpha\right]}{\delta A_h^l[z]}\right\}+\mathcal{O}\left(\theta^2\right).
\label{Application_EtoU}
\end{eqnarray}
The right hand side of ($\ref{Application_EtoU}$) has now to be calculated by using the relation ($\ref{relation_dUdA}$)
for the application of the variation with respect to the connection to the holonomy and simplifying
the obtained expression. This leads to

\begin{eqnarray}
\hat E_i(\mathcal{S})U\left[A,\alpha\right]&=&-G i \int_{\mathcal{S}}d \sigma^1 d \sigma^2 \int_\gamma ds\
\epsilon_{abc}\frac{\partial x^a(\vec \sigma)}{\partial \sigma_1}
\frac{\partial x^b(\vec \sigma)}{\partial \sigma_2}
\left\{8\pi\beta\ \delta\left(x(\vec \sigma)-\gamma(s)\right)
\frac{d \gamma^c(s)}{ds}U\left[A,\gamma_1\right]\tau_i U\left[A,\gamma_2\right]
\right.\nonumber\\&&\left.
+i\theta^{mn}\delta^{kp}\delta^{lq}\delta_{wz}\delta_{ir}\mathcal{X}^{ezcr}_{gphq}
\partial_m A_e^w[x(\vec \sigma)]\int d^3 z
\left[\partial_n \delta\left(z-x(\vec \sigma)\right)
\right.\right.\nonumber\\&&\left.\left.
\times \frac{\delta}{\delta A_g^k[z]}
\left(\delta\left(z-\gamma(s)\right)\frac{d \gamma^h(s)}{ds}U\left[A,\gamma_1\right]\tau_l U\left[A,\gamma_2\right]\right)\right]\right\}
+\mathcal{O}\left(\theta^2\right)\nonumber\\
&=&-G i \int_{\mathcal{S}}d \sigma^1 d \sigma^2 \int_\gamma ds\
\epsilon_{abc}\frac{\partial x^a(\vec \sigma)}{\partial \sigma_1}
\frac{\partial x^b(\vec \sigma)}{\partial \sigma_2}
\left\{8\pi\beta\ \delta\left(x(\vec \sigma)-\gamma(s)\right)
\frac{d \gamma^c(s)}{ds} U\left[A,\gamma_1\right]\tau_i U\left[A,\gamma_2\right]
\right.\nonumber\\&&\left.
+iG\theta^{mn}\delta^{kp}\delta^{lq}\delta_{wz}\delta_{ir}\mathcal{X}^{ezcr}_{gphq}\partial_m A_e^w[x(\vec \sigma)]
\int d^3 z \left[\partial_n \delta\left(z-x(\vec \sigma)\right)
\right.\right.\nonumber\\&&\left.\left.
\times \delta\left(z-\gamma(s)\right)\frac{d \gamma^h(s)}{ds}\int_{\gamma_1}
ds^{\prime} \frac{d \gamma^g_1(s^{\prime})}{ds^{\prime}}
\delta\left(z-\gamma_1(s^{\prime})\right)
U\left[A,\gamma_1\right]\tau_k\tau_l U\left[A,\gamma_2\right]\right]
\right\}+\mathcal{O}\left(\theta^2\right)\nonumber\\
&=&-G i \int_{\mathcal{S}}d \sigma^1 d \sigma^2 \int_\gamma ds\
\epsilon_{abc}\frac{\partial x^a(\vec \sigma)}{\partial \sigma_1}
\frac{\partial x^b(\vec \sigma)}{\partial \sigma_2}
\left\{8\pi\beta\ \delta\left(x(\vec \sigma)-\gamma(s)\right)
\frac{d \gamma^c(s)}{ds} U\left[A,\gamma_1\right]\tau_i U\left[A,\gamma_2\right]
\right.\nonumber\\&&\left.
-iG\theta^{mn}\delta^{kp}\delta^{lq}\delta_{wz}\delta_{ir}\mathcal{X}^{ezcr}_{gphq}\partial_m A_e^w[x(\vec \sigma)]
\partial_n \left(\delta\left(x(\vec \sigma)-\gamma(s)\right)\frac{d \gamma^h(s)}{ds}
\right.\right.\nonumber\\&&\left.\left.
\times \int_{\gamma_1}ds^{\prime} \frac{d \gamma^g_1(s^{\prime})}{ds^{\prime}}
\delta\left(x(\vec \sigma)-\gamma_1(s^{\prime})\right)
U\left[A,\gamma_1\right]\tau_k\tau_l U\left[A,\gamma_2\right]\right)
\right\}+\mathcal{O}\left(\theta^2\right).
\label{calculation_application_E_U}
\end{eqnarray}
In ($\ref{calculation_application_E_U}$) with respect to the second application of the variation with respect
to the connection has been used the assumption that the intersection point $p$ belongs to the first section
of the devision of $\gamma$, $\gamma_1$ namely. In the last step has in the term depending on the noncommutativity
parameter $\theta$ been used partial integration with respect to $z$ and the fact that because of the delta
function the expression within the integral vanishes at the boundary of the integral. Since the derivative
does not act then on the delta function anymore, but on the other terms the integral refers to, the $z$-integral breaks
down because of the delta function. The first expression in the bracket of the last step of ($\ref{calculation_application_E_U}$) corresponds to the usual expression without generalization,
which is well-known in the literature, see \cite{Rovelli:2004} or \cite{Kiefer:2004} for example,
and can as usual be determined to:

\begin{eqnarray}
&&-8\pi\beta G i \int_{\mathcal{S}}d \sigma^1 d \sigma^2 \int_\gamma ds\
\epsilon_{abc}\frac{\partial x^a(\vec \sigma)}{\partial \sigma_1}
\frac{\partial x^b(\vec \sigma)}{\partial \sigma_2}\left\{ 
\delta\left(x(\vec \sigma)-\gamma(s)\right)\left[\frac{d \gamma^c(s)}{ds} U\left[A,\gamma_1\right]\tau_i U\left[A,\gamma_2\right]\right]\right\}\nonumber\\&&=\pm 8\pi \beta G i U\left[A,\gamma_1\right]\tau_i U\left[A,\gamma_2\right],
\label{application_E_U_usual}
\end{eqnarray}
where has to be used the following substitution of variables: $(\sigma^1,\sigma^2,s)\longrightarrow (x^1,x^2,x^3)$
and accordingly the relation:

\begin{equation}
\int_{\mathcal{S}} d \sigma^1 d \sigma^2 \int_\gamma ds\
\epsilon_{abc}\frac{\partial x^a(\vec \sigma)}{\partial \sigma_1}
\frac{\partial x^b(\vec \sigma)}{\partial \sigma_2}\frac{d \gamma^c(s)}{ds}\delta\left(x(\vec \sigma)-\gamma(s)\right)
=\int dx^1 dx^2 dx^3 \delta\left(x(\vec \sigma)-\gamma(s)\right)=\pm 1.
\label{integral_substitution}
\end{equation}
The second expression of course arises from the special assumption of a noncommutative space-time and is accordingly
characteristic for the generalization of this paper. The calculation of this second expression
is much more intricated,

\begin{eqnarray}
&&-G^2 \int_{\mathcal{S}}d \sigma^1 d \sigma^2 \int_\gamma ds\
\epsilon_{abc}\frac{\partial x^a(\vec \sigma)}{\partial \sigma_1}
\frac{\partial x^b(\vec \sigma)}{\partial \sigma_2}
\left\{\theta^{mn}\delta^{kp}\delta^{lq}\delta_{wz}\delta_{ir}\mathcal{X}^{ezcr}_{gphq}\partial_m A_e^w[x(\vec \sigma)]
\right.\nonumber\\&&\left.\quad\quad\quad\quad\quad
\times \partial_n \left(\delta\left(x(\vec \sigma)-\gamma(s)\right)\frac{d \gamma^h(s)}{ds}\int_{\gamma_1}
ds^{\prime} \frac{d \gamma^g_1(s^{\prime})}{ds^{\prime}}
\delta\left(x(\vec \sigma)-\gamma_1(s^{\prime})\right)
U\left[A,\gamma_1\right]\tau_k\tau_l U\left[A,\gamma_2\right]\right)
\right\}+\mathcal{O}\left(\theta^2\right)
\nonumber\\
&&=G^2 \int_{\mathcal{S}}d \sigma^1 d \sigma^2 \int_\gamma ds\
\int_{\gamma_1} ds^{\prime}\
\epsilon_{abc}\frac{\partial x^a(\vec \sigma)}{\partial \sigma_1}
\frac{\partial x^b(\vec \sigma)}{\partial \sigma_2}\left\{\delta\left(x(\vec \sigma)-\gamma_1(s^{\prime})\right)
\right.\nonumber\\&&\left.\quad\quad\quad\quad\quad\times
\theta^{mn}\delta^{kp}\delta^{lq}\delta_{wz}\delta_{ir}\mathcal{X}^{ezcr}_{gphq}
\partial_m \partial_n A_e^w[x(\vec \sigma)]
\left(\frac{d \gamma^h(s)}{ds}\frac{d \gamma^g_1(s^{\prime})}{ds^{\prime}}
\delta\left(x(\vec \sigma)-\gamma(s)\right)
U\left[A,\gamma_1\right]\tau_k\tau_l U\left[A,\gamma_2\right]\right)
\right\}+\mathcal{O}\left(\theta^2\right)
\nonumber\\
&&=G^2 \int_{\mathcal{S}}d \sigma^1 d \sigma^2 \int_\gamma ds\
\int_{\gamma_1} ds^{\prime}\
\epsilon_{abc}\frac{\partial x^a(\vec \sigma)}{\partial \sigma_1}
\frac{\partial x^b(\vec \sigma)}{\partial \sigma_2}
\frac{d \gamma^c_1(s^{\prime})}{ds^{\prime}}
\left\{\delta\left(x(\vec \sigma)-\gamma_1(s^{\prime})\right)
\right.\nonumber\\&&\left.\quad\quad\quad\quad\quad\times
\theta^{mn}\partial_m \partial_n A_h^l[x(\vec \sigma)]
\left(\frac{d \gamma^h(s)}{ds}
\delta\left(x(\vec \sigma)-\gamma(s)\right)
U\left[A,\gamma_1\right]\tau_i\tau_l U\left[A,\gamma_2\right]\right)
\right\}+\mathcal{O}\left(\theta^2\right)
\nonumber\\
&&=G^2 \int_\gamma ds\ \frac{d \gamma^h(s)}{ds}
\left\{\theta^{mn}\partial_m \partial_n A_h^l[p]
\delta\left(p-\gamma(s)\right) U\left[A,\gamma_1\right]\tau_i\tau_l U\left[A,\gamma_2\right]
\right\}+\mathcal{O}\left(\theta^2\right)
\nonumber\\
&&=G^2 \int_\gamma d \gamma^h \left\{\theta^{mn}\partial_m \partial_n A_h^l[p]
\delta\left(p-\gamma\right) U\left[A,\gamma_1\right]\tau_i\tau_l U\left[A,\gamma_2\right]
\right\}+\mathcal{O}\left(\theta^2\right)
\nonumber\\
&&=\pm G^2 \theta^{mn}\sum_h \partial_m \partial_n A_h^l[p]
U\left[A,\gamma_1\right]\tau_i\tau_l U\left[A,\gamma_2\right]
+\mathcal{O}\left(\theta^2\right).
\label{calculation_application_E_U_generalization}
\end{eqnarray}
In the first step of ($\ref{calculation_application_E_U_generalization}$) has again been used partial
integration and the fact that because of the delta function the expression within the integral vanishes at
the boundary of the integral and the expression has been reordered with respect to its factors.
After this, in the second step, has been applied the definition of the components $\mathcal{X}^{abcd}_{efgh}$
of the tensor $\mathcal{X}$, which are given in ($\ref{definition_X}$), and the Kronecker deltas have been contracted.
Then, in the third step, three of the integrals have been solved in analogy to ($\ref{application_E_U_usual}$)
by applying ($\ref{integral_substitution}$) with $s$ replaced by $s^{\prime}$ and another reordering has been
taken place. Finally, in the last two steps, the remaining integral has been solved. This means that the
application of the smeared out version of the canonical conjugated operator to the holonomy of the
connection is given by

\begin{eqnarray}
\hat E_i(\mathcal{S})U\left[A,\alpha\right]&=&\pm G i
\left(8\pi \beta\ U\left[A,\gamma_1\right]\tau_i U\left[A,\gamma_2\right]
+iG\theta^{mn}\sum_a \partial_m \partial_n A_a^j[p]
U\left[A,\gamma_1\right]\tau_i\tau_j
U\left[A,\gamma_2\right]\right)+\mathcal{O}\left(\theta^2\right).
\label{relation_EU}
\end{eqnarray}
To obtain a gauge invariant expression, the operator ($\ref{smeared_area}$) has to be considered
quadratically and thus it has to be applied another time to the holonomy. This means nothing else
but that the relation ($\ref{relation_EU}$) is iterated and yields

\begin{eqnarray}
\hat E^2(\mathcal{S})U\left[A,\alpha\right]&=&\pm G i\ \hat E(\mathcal{S})
\left(8\pi \beta\ U\left[A,\gamma_1\right]\tau_i U\left[A,\gamma_2\right]
+iG\theta^{mn}\sum_a \partial_m \partial_n A_a^j[p] U\left[A,\gamma_1\right]
\tau_i\tau_j U\left[A,\gamma_2\right]\right)+\mathcal{O}\left(\theta^2\right),
\nonumber\\
&=&-8\pi\beta G^2\sum_i \left(8\pi\beta\ U\left[A,\gamma_1\right]\tau_i \tau_i U\left[A,\gamma_2\right]
+iG\theta^{mn}\sum_a \partial_m \partial_n A_a^j[p] U\left[A,\gamma_1\right]
\tau_i\tau_j\tau_i U\left[A,\gamma_2\right]
\right.\nonumber\\&&\left.\quad\quad\quad\quad\quad
+iG\theta^{mn}\sum_a \partial_m \partial_n A_a^j[p]U\left[A,\gamma_1\right]
\tau_i\tau_i\tau_j U\left[A,\gamma_2\right]\right)
+\mathcal{O}\left(\theta^2\right),
\end{eqnarray}
where has been used that the second application to the operator ($\ref{smeared_area}$) only acts on
the holonomy referring to the first section of the devision of $\gamma$, since the point $P$ belongs
to this section, $\gamma_1$ namely. A spin network state is defined as

\begin{equation}
|\Psi_S[A]\rangle=R^{I}_{\alpha\beta}(U[\alpha,\gamma])\Psi^{\alpha\beta}[A],
\end{equation}
where the transformation operator $R^{I}_{\alpha\beta}(U[\alpha,\gamma])$ describes the holonomy along the curve $\gamma$
in the irreducible representation corresponding to the spin $I$. It is assumed that the spin network $S$ intersects
the surface $\mathcal{S}$ at a single point $p$. The application of the square of the smeared out operator
($\ref{smeared_area}$) to a spin network state leads to

\begin{eqnarray}
\hat E^2(\mathcal{S})|\Psi_S[A]\rangle&=&
8\pi\beta G^2
\left\{8\pi\beta\ I(I+1)+iG\theta^{mn}\sum_a \partial_m \partial_n A_a^j[p]
\left[I(I+1)\tau_j+\sum_i \lambda_{ijk}\tau_k\tau_i\right]
\right.\nonumber\\&&\left.\quad\quad\quad\quad\quad
+iG\theta^{mn}\sum_a \partial_m \partial_n A_a^j[p]\ I(I+1)\tau_j\right\}|\Psi_S[A]\rangle
\nonumber\\
&=&8\pi\beta G^2 \left\{8\pi\beta I(I+1)
+iG \theta^{mn}\sum_a \partial_m \partial_n A_d^j[p]\left[2I(I+1)\tau_j
+\sum_i \lambda_{ijk}\tau_k\tau_i\right]\right\}|\Psi_S[A]\rangle,
\end{eqnarray}
where have been used the following properties of the generators $\tau_i$:

\begin{equation}
\sum_i \tau_i^{(I)} \tau_i^{(I)}=-I(I+1)\quad,\quad \left[\tau_i^{(I)},\tau_j^{(I)}\right]
=i\lambda_{ijk}^{(I)}\tau_k^{(I)} \Leftrightarrow 
\tau_i^{(I)} \tau_j^{(I)}=\tau_j^{(I)} \tau_i^{(I)}+i\lambda_{ijk}^{(I)}\tau_k^{(I)},
\label{relation_generators}
\end{equation}
where the $\lambda_{ijk}^{(I)}$ denote the structure constants of the Lie Algebra describing the symmetry group
defining the space of the spin $I$. The first relation of ($\ref{relation_generators}$) is also used in the usual case,
whereas the Lie Algebra is only needed with respect to the generalization term. If there is assumed that the spin network $S$
intersects $\mathcal{S}$ at a finite number of points $N$, then the expression of the area operator
applied to a spin network state is given by
 
\begin{eqnarray}
\hat {\mathcal{A}}(\mathcal{S})|\Psi_S[A]\rangle&=&\sum_{N}\sqrt{\hat E^2(\mathcal{S}_N)}|\Psi_S[A]\rangle\nonumber\\
&=&\sqrt{8\pi\beta}\ G \sum_{P \in S \cap \mathcal{S}}
\sqrt{8\pi\beta\ I(I+1)+iG \theta^{mn}\sum_a \partial_m \partial_n A_a^j[p]\left[2I(I+1)\tau_j
+\sum_i \lambda_{ijk}\tau_k\tau_i\right]}|\Psi_S[A]\rangle.
\end{eqnarray}
This means nothing else, but that the generalized version of the area operator in loop quantum gravity
on noncommutative space-time has been determined by considering a calculation to the first order in the
noncommutativity parameter $\theta$. The calculation has been performed in complete analogy to the
usual case by using the generalized representation of the smeared out version of the canonical conjugated
operator to the connection operator according to the generalization of the quantization of the 
gravitational field on noncommutative space-time.

\section{Discussion}

In the present paper canonical quantum gravity on noncommutative space-time has been considered.
This theory represents a combination of two very important concepts in contemporary fundamental
theoretical physics, the canonical quantization of general relativity on the one hand and
noncommutative geometry on the other hand.
The presented theory is based on Hamiltonian gravity on noncommutative space-time formulated by
using the moyal star product to represent the expressions depending on noncommuting quantities to
quantities on usual space-time. Although the transition to the corresponding quantum theory
is performed as usual by postulating canonical commutation relations between the
quantity describing the gravitational field and its canonical conjugated quantity,
the quantization condition has to be generalized because of the necessity to generalize
commutators on a noncommutative space-time. After the formulation of the canonical quantum
theory of quantum geometrodynamics on noncommutative space-time under incorporation of
a coupling to a matter field, the Ashtekar formalism has been considered. In both cases
the generalized expressions of the representations of the operators with respect to the field
and the corresponding quantum constraints have been given. To formulate the representation of
the operator describing the canonical conjugated quantity, it was necessary to introduce a special
tensor of eight order.
At the end, the holonomy representation corresponding to the connection in the Ashtekar formalism,
on which loop quantum gravity is based on, has been used to calculate the generalized area operator in case
of noncommutative geometry. This calculation has been performed to the first order in the noncommutativity
parameter $\theta$ as all calculations in this paper. In this approximation the generalized area operator
in loop quantum gravity on noncommutative space-time has finally been determined.
The generalized quantization principle as basic constituent of the presented theory, which is
implied by the direct combination of the assumption of noncommutative geometry and field
quantization, is a different version of a generalized quantization principle as it has been
considered with respect to the variables of quantum mechanics as generalized uncertainty principle \cite{Maggiore:1993kv},\cite{Kempf:1994su},\cite{Hinrichsen:1995mf} and in quantum gravity 
\cite{Kober:2011uj},\cite{Kober:2015bkv},\cite{Kober:2014xxa},\cite{Majumder:2011ad},\cite{Majumder:2011bv}.
Usually, noncommutative geometry is mainly interpreted as a possible extension of the structure
of space-time, which could possibly cure the appearance of divergencies in quantum field theory, since
the presupposed noncommutative geometry implies the existence of a minimal length.
This is in particular very interesting with respect to the attempt of a description of quantum gravity
in the framework of usual quantum field theory because of its nonrenormalizability, which could be
omitted by the appearing minimal length. However, if this would be true,
then it has in an analogous way to be possible, to give a corresponding canonical formulation of
quantum gravity on noncommutative space-time. This has been done in this paper. It remains an open
task and would lead to very interesting projects for further research to consider extended theories
like canonical quantum supergravity on noncommutative space-time.

\end{document}